\documentclass[12pt]{article}
\usepackage{amsmath,amssymb,amsthm}
\usepackage{latexsym,graphicx,color,bbm,subfigure}

\newcommand{\NF}{N_{\rm f}}

\newcommand{\beq}{\begin{eqnarray}}
\newcommand{\eeq}{\end{eqnarray}}
\newcommand{\non}{\nonumber\\}
\newcommand{\D}{\mathcal{D}}
\newcommand{\p}{\partial}
\newcommand{\Tr}{{\rm Tr}}
\newcommand{\diag}{{\rm diag}}
\newcommand{\sign}{{\rm sign}}

\makeatletter
\@addtoreset{equation}{section}

\renewcommand{\thefootnote}{\fnsymbol{footnote}}
\newcommand{\thetablename}{Table}
\def\fnum@table{\thetablename\ \thetable}
\makeatother

\begin{document}

\thispagestyle{empty}
\begin{flushright}
IFUP-TH/2009-12
\end{flushright}
\vspace{10mm}

\begin{center}
{\Large \bf Non-Abelian Chern-Simons vortices\\ with generic gauge groups} 
\\[15mm]
{Sven Bjarke~{\sc Gudnason}}\footnote{\it e-mail address:
gudnason(at)df(dot)unipi(dot)it}
\vskip 6 mm

\bigskip\bigskip
{\it
Department of Physics, University of Pisa, \\
INFN, Sezione di Pisa,\\
Largo Pontecorvo, 3, Ed. C, 56127 Pisa, Italy
}

\bigskip

\bigskip

{\bf Abstract}\\[5mm]
{\parbox{14cm}{\hspace{5mm}
\small
We study non-Abelian Chern-Simon BPS-saturated vortices enjoying
${\cal N}=2$ supersymmetry in $d=2+1$ dimensions, with generic gauge
groups of the form $U(1)\times G'$, with $G'$ being a simple group,
allowing for orientational modes in the solutions. We will keep the
group as general as possible and utilizing the powerful moduli matrix
formalism to provide the moduli spaces of vortices and
derive the corresponding master equations. Furthermore, we study
numerically the vortices applying a radial Ansatz to solve the
obtained master equations and we find especially a splitting of the
magnetic fields, when the coupling constants for the trace-part
and the traceless part of the Chern-Simons term are varied, such that
the Abelian magnetic field density can become negative near the origin
of the vortex while the non-Abelian part stays positive, and vice
versa.  
}}
\end{center}
\newpage
\pagenumbering{arabic}
\setcounter{page}{1}
\setcounter{footnote}{0}
\renewcommand{\thefootnote}{\arabic{footnote}}

\section{Introduction}

Solitons play a crucial role in a vast area of physics ranging from
particle physics and cosmology to condensed matter physics. Planar
physics i.e.~in 2+1 dimensions, has radically different properties as
the spin is not quantized as in 3+1 dimensions giving rise to the
anyons among others, objects having fractional spin and
statistics. This 
can be realized by the Chern-Simons term which has been widely used in
e.g.~the theory of the fractional quantum Hall effect
\cite{Zhang:1988wy}. 
Another aspect is that the high temperature limit of a four
dimensional theory can be described by a three dimensional one, where
the Chern-Simons term resides naturally. Another interesting feature
of Chern-Simons theories is that it provides a gauge invariant
mechanism of mass generation \cite{Deser:1981wh}.

The most celebrated vortex solution, namely the
Abrikosov-Nielsen-Olesen (ANO) vortex was found half a century ago
\cite{Abrikosov:1956sx,Nielsen:1973cs}. This object carries magnetic
flux in its interior. Later, similar vortex solutions, however in 2+1
dimensions where found with a Chern-Simons term instead of a Maxwell
term \cite{Hong:1990yh,Jackiw:1990pr}. These vortices possess the
already mentioned features of fractional spin and statistics,
viz.~they are anyon-like. Furthermore, there exist vortices
in both the asymmetric phase (like for the ANO vortices) and also in
the symmetric phase. The latter do not have a topological
argument for stability. The vortices with Maxwell or with Chern-Simons
terms split into three categories depending on the self-coupling of
the Higgs field, viz.~type I/II vortices or the critical BPS saturated
vortices \cite{Bogomolny:1975de}, where the vortices attract, repel
and do not feel any force among the selves, respectively. The latter
corresponds to some amount of supersymmetry being present in the
theories at hand. Recently, a fourth type of vortices in the Abelian
Chern-Simons model has been found, behaving as a type I vortex at small
amount of magnetic field and turns into type II when the magnetic
field piles up repelling further vortices from the clusters
\cite{Bolognesi:2007ez}. This type of vortex was denoted a type III
vortex. 

A few years ago, non-Abelian vortices have been discovered
\cite{Hanany:2003hp,Auzzi:2003fs}, being flux tubes which are carrying
orientational modes. These models have been extensively studied with
the gauge group
$U(N)\simeq U(1)\times SU(N)/\mathbb{Z}_N$ and only recently with
generalizations to other groups
\cite{Ferretti:2007rp,Eto:2008yi,Eto:2009bg}. In particular the moduli
space of these vortices have been studied in detail 
\cite{Hanany:2003hp,Auzzi:2003fs,Eto:2009bg,Eto:2005yh,Auzzi:2005gr,Eto:2006cx}. 
Good reviews summarizing many results can be found in
Refs.~\cite{Tong:2005un,Eto:2006pg,Shifman:2007ce}. 

The first studies of non-Abelian Chern-Simons vortices are made with 
a simple group, viz.~$SU(2)$ and $SU(N)$ with fields in the adjoint
representation \cite{deVega:1986hm,Kumar:1986yz,deVega:1986eu} and
later numerical solutions have been found
\cite{NavarroLerida:2008uj}. In
Refs.~\cite{Aldrovandi:2007nb,Lozano:2007yz} the 
non-Abelian Chern-Simons vortices have been studied with a $U(N)$
gauge group allowing for orientational modes to be present and they
identified the moduli space of a single vortex solution. 
Furthermore Refs.~\cite{Collie:2008mx,Collie:2008za,Buck:2009pd} have
considered packaging together the Yang-Mills and the non-Abelian
Chern-Simons terms for $U(N)$ gauge groups. In
Ref.~\cite{Collie:2008mx} the dynamics of the vortices has been
studied and in Ref.~\cite{Collie:2008za} in addition to the
topological charge, conserved Noether charges associated with a
$U(1)^{N-1}$ flavor symmetry of the theory due to inclusion of a mass
term for the squarks. In Ref.~\cite{Buck:2009pd} 
numerical solutions have been provided. 

Many related topics can be found in the excellent reviews
\cite{Dunne:1998qy,Horvathy:2008hd}.

It is the purpose of this paper to consider a wider class of
non-Abelian Chern-Simons vortices carrying orientational modes, with
the gauge group kept as general as possible, except when we will do
some concrete numerical calculations.  

\section{The model}
Our starting point will be the Yang-Mills-Chern-Simons-Higgs theory. 
We are considering the following ${\cal N}=2$ supersymmetric theory
(viz.~with $4$ supercharges) in $d=2+1$ dimensions with the gauge group
$G=U(1)\times G'$, where $G'$ is a simple group. 
The bosonic part of the Lagrangian density reads
\begin{align}
\mathcal{L}_{\rm YMCSH} = &
-\frac{1}{4g^2}\left(F_{\mu\nu}^a\right)^2
-\frac{1}{4e^2}\left(F_{\mu\nu}^0\right)^2
-\frac{\mu}{8\pi}\epsilon^{\mu\nu\rho}
  \left(A_{\mu}^a\partial_\nu A_\rho^a  
  -\frac{1}{3}f^{abc}A_{\mu}^a A_{\nu}^b A_{\rho}^c\right)
-\frac{\kappa}{8\pi}
  \epsilon^{\mu\nu\rho} A_{\mu}^0\partial_\nu A_\rho^0 \non
&+\frac{1}{2g^2}\left(\D_\mu\phi^a\right)^2
+\frac{1}{2e^2}\left(\partial_\mu\phi^0\right)^2
+\Tr\left(\D_\mu H\right)\left(\D^\mu H\right)^\dag
-\Tr \left|\phi H - H m\right|^2\non
&-\frac{g^2}{2}\left(\Tr\left(HH^\dag t^a\right)-\frac{\mu}{4\pi}\phi^a\right)^2
-\frac{e^2}{2}\left(\Tr\left(HH^\dag t^0\right)-\frac{\kappa}{4\pi}\phi^0
  -\frac{1}{\sqrt{2N}}\xi\right)^2 \ , \label{eq:LYMCSH}
\end{align}
where $a=1,\ldots,\dim(G')$, the index $0$ is for the
Abelian group and $\alpha=0,1,\ldots,\dim(G')$ and we use
the conventions 
\begin{align}
F_{\mu\nu} = \p_\mu A_\nu - \p_\nu A_\mu + i\left[A_\mu,A_\nu\right] \ ,
\quad
\D_\mu H = \left(\p_\mu + i A_\mu\right)H \ , 
\quad
\D_\mu\phi &= \p_\mu\phi + i \left[A_\mu,\phi\right]
\ .  
\end{align}
$A_\mu = A_\mu^\alpha t^\alpha$ is the gauge potential, $F_{\mu\nu}$
is the field strength, $\phi$ is an adjoint scalar field which we can
take to be real and finally $H$ is a color-flavor matrix of dimension
$N \times \NF$ of $\NF$ matter fields.
We will define $N\equiv\dim(R_{G'})$ but for simplicity we choose the 
representation $R_{G'}$ as the fundamental one of $G'$. 
We are using the following normalization of the generators
\beq t^0 = \frac{\mathbf{1}_N}{\sqrt{2N}} \ , \quad
\Tr\left(t^a t^b\right) = \frac{1}{2}\delta^{a b} \ . \eeq
There are four coupling constants entering our game at this point;
$e\in\mathbb{R}$ is the Abelian coupling of the Yang-Mills kinetic
term (Maxwell), $g\in\mathbb{R}$ the is the coupling for the
semi-simple part of the Yang-Mills kinetic term, which corresponds to
$G'$. $\kappa\in\mathbb{R}$ is the Abelian coupling of the
Chern-Simons term while $\mu\in\mathbb{Z}$ are solely integers to
render the non-Abelian Chern-Simons action gauge invariant up to
large gauge transformations \cite{Deser:1982vy}. $\xi$ is a
Fayet-Iliopoulos parameter. Finally, $m$ is a mass matrix which we
will set to zero in this paper.

The scope of study in this paper will be on the Chern-Simons part of
this theory.
A detailed study of the vortices dependence of the parameters of the
model above with also the Yang-Mills term in action will be done
elsewhere \cite{MESBG}.

\section{Non--Abelian Chern--Simons--Higgs theory}

Now let us take the limit $e\to\infty,g\to\infty, m=0$ and
$\kappa\neq\mu$ and in turn integrate out the adjoint scalar field
$\phi$:
\begin{align}
\phi^a = \frac{4\pi}{\mu}\Tr\left(H H^\dag t^a\right) \ , \quad
\phi^0 = \frac{4\pi}{\kappa}\frac{1}{\sqrt{2N}}\left[\Tr\left(H H^\dag\right) 
  -\xi\right] \ .
\end{align}
This leaves us with the non-Abelian Chern-Simons theory
\begin{align}
\mathcal{L}_{\rm CSH} &=
-\frac{\mu}{8\pi}\epsilon^{\mu\nu\rho}
  \left(A_{\mu}^a\partial_\nu A_\rho^a  
  -\frac{1}{3}f^{abc}A_{\mu}^a A_{\nu}^b A_{\rho}^c\right)
-\frac{\kappa}{8\pi}\epsilon^{\mu\nu\rho}
  \left(A_{\mu}^0\p_\nu A_\rho^0\right)
+\Tr\left(\D_\mu H\right)^\dag\left(\D^\mu H\right) \non&\phantom{=\ }
-4\pi^2\Tr\left|\left\{
  \frac{\mathbf{1}_N}{N\kappa}\left(\Tr\left(H H^\dag\right)
  -\xi\right)
  +\frac{2}{\mu}\Tr\left(H H^\dag t^a\right)t^a\right\} H\right|^2
  \ ,
\end{align}
which will be the main focus of this paper.
It still enjoys
${\cal N}=2$ supersymmetry and there are 3 parameters governing the
solutions; the Abelian Chern-Simons coupling $\kappa$ and the
non-Abelian Chern-Simons coupling $\mu$ and finally the winding 
number $\nu=\frac{k}{n_0}$ \cite{Eto:2008yi}. $n_0$ denotes the
greatest common divisor (gcd) of the Abelian charges of the
holomorphic invariants of $G'$, see \cite{Eto:2008yi}. For simple
groups this coincides with the center as $\mathbb{Z}_{n_0}$. We will
take $k>0$. 

There are three different phases of the theory at hand. An unbroken
phase with $\langle H\rangle=0$ and a broken phase with $\langle
H\rangle=\sqrt{\frac{\xi}{N}}$. In between there are partially broken
phases. We will consider only the completely broken phase in this
paper. 

The equations of motion are
\begin{align}
\frac{\mu}{8\pi}\epsilon^{\mu\nu\sigma} F_{\mu\nu}^a &=
-i\Tr\left[H^\dag t^a \D^\sigma H - \left(\D^\sigma H\right)^\dag
  t^a H \right] \ , \label{eq:GausslawNA}\\
\frac{\kappa}{8\pi}\epsilon^{\mu\nu\sigma} F_{\mu\nu}^0 &=
-i\Tr\left[H^\dag t^0 \D^\sigma H - \left(\D^\sigma H\right)^\dag
  t^0 H \right] \ ,\label{eq:GausslawA}
\end{align}
\begin{align}
\D_\mu\D^\mu H &=
-4\pi^2\left[
  \frac{\mathbf{1}_N}{N\kappa}\left(\Tr\left(H H^\dag\right) 
  -\xi\right)
  +\frac{2}{\mu}\Tr\left(H H^\dag t^a\right)t^a\right]^2 H 
  \non&\phantom{=\ }
-\frac{8\pi^2}{N\kappa}\Tr\left(\left[
  \frac{\mathbf{1}_N}{N\kappa}\left(\Tr\left(H H^\dag\right) 
  -\xi\right)
  +\frac{2}{\mu}\Tr\left(H H^\dag t^a\right)t^a\right] H H^\dag\right)
  H \non &\phantom{=\ }
-\frac{16\pi^2}{\mu}\Tr\left(\left[
  \frac{\mathbf{1}_N}{N\kappa}\left(\Tr\left(H H^\dag\right) 
  -\xi\right)
  +\frac{2}{\mu}\Tr\left(H H^\dag t^b\right)t^b\right] 
  H H^\dag t^a\right) t^a H \ .
\end{align}
The tension, defined by the integral on the plane over the time-time
component of the energy-momentum tensor, is given by
\begin{align}
T =&\ \int_{\mathbb{C}}\Tr\left\{\left|\D_0 H\right|^2
+\left|\D_i H\right|^2 
+4\pi^2\left|\left(
  \frac{\mathbf{1}_N}{N\kappa}\left(\Tr\left(H H^\dag\right)
  -\xi\right)
  +\frac{2}{\mu}\Tr\left(H H^\dag t^a\right)t^a\right)
  H\right|^2\right\} \ ,
\end{align}
which by a standard Bogomol'nyi completion can be rewritten as
\begin{align}
T =&\ \int_{\mathbb{C}}\Tr\bigg\{
\left|\D_0 H - i2\pi
  \left(
  \frac{\mathbf{1}_N}{N\kappa}\left(\Tr\left(H H^\dag\right)
  -\xi\right)
  +\frac{2}{\mu}\Tr\left(H H^\dag t^a\right)t^a\right)
  H\right|^2
+ 4\left|\bar{\D}H\right|^2\bigg\} \non&\ 
-\frac{\xi}{\sqrt{2N}}\int_{\mathbb{C}}F_{12}^0 
+i\Tr\int_{\mathbb{C}}\left[\p_2\left(H^\dag\D_1 H\right)
-\p_1\left(H^\dag\D_2 H\right)\right] \ .
\end{align}
This leads immediately to the BPS-equations which need to be
accompanied by the Gauss law being the $\sigma=0$ component of the
Eqs.~(\ref{eq:GausslawNA}),(\ref{eq:GausslawA}) 
\begin{align}
\bar{\D}H = 0 \ , \quad
\D_0 H = i2\pi
  \left(
  \frac{\mathbf{1}_N}{N\kappa}\left(\Tr\left(H H^\dag\right)
  -\xi\right)
  +\frac{2}{\mu}\Tr\left(H H^\dag t^a\right)t^a\right)
  H \ .
\end{align}
Rewriting the boundary term using the first BPS-equation, we have for
the BPS saturated vortices the tension
\beq T = - \frac{\xi}{\sqrt{2N}}\int_{\mathbb{C}}F_{12}^0 
+\frac{1}{2}\Tr\int_{\mathbb{C}} \p_i^2\left(H H^\dag\right) 
= 2\pi\xi\nu \ , \label{eq:energydensity1} \eeq
with $\nu$ being the $U(1)$ winding number.
By combining the BPS equations with the Gauss law, we obtain the
following system 
\begin{align}
\bar{\D}H &= 0 \ , \\
F_{12}^a t^a &= \frac{16\pi^2}{N\kappa\mu}
  \left(\Tr\left(H H^\dag\right) -\xi\right)
  \Tr\left(H H^\dag t^a\right)t^a
+\frac{16\pi^2}{\mu^2}\Tr\left(H H^\dag t^b\right)
  \Tr\left(H H^\dag\left\{t^a,t^b\right\}\right)t^a \ , \\
F_{12}^0 t^0 &= \frac{8\pi^2}{N^2\kappa^2}\Tr\left(H H^\dag\right)
  \left(\Tr\left(H H^\dag\right)-\xi\right)\mathbf{1}_N 
+\frac{16\pi^2}{N\kappa\mu}
  \left(\Tr\left(H H^\dag t^a\right)\right)^2\mathbf{1}_N \ .
\label{eq:NACSmastersystem}
\end{align}
An interesting comment is that the system only depends on three
combinations of the couplings; viz.~$\kappa^2$, $\mu^2$ and
$\kappa\mu$. There are thus only two choices of signs giving different
solutions $\sign(\kappa)=\pm\sign(\mu)$.
This system is of a generic character and one can readily
apply one's favorite group. 
Setting $\kappa=\mu$, the BPS-equations become 
\begin{align}
\bar{\D} H = 0 \ , \quad
\D_0 H = \frac{i2\pi}{\kappa}
  \left[2\Tr\left(H H^\dag t^\alpha\right)t^\alpha -
  \frac{\xi}{N}\mathbf{1}_N\right] H \ ,
\end{align}
which in turn yields the simplified system by combination with the
Gauss law
\begin{align}
\bar{\D} H = 0 \ , \quad
F_{12}^\alpha t^\alpha =
\frac{16\pi^2}{\kappa^2}\left[
\Tr\left(H H^\dag\left\{t^\alpha,t^\beta\right\}\right)
\Tr\left(H H^\dag t^\beta\right)
-\frac{\xi}{N}\Tr\left(H H^\dag t^\alpha\right)\right] t^\alpha \ .
\end{align}
In the next section, we will consider the
cases of $G'=SU(N)$, $G'=SO(N)$ and $G'=USp(2M)$, and finally make the
corresponding master equations.

\subsection{Master equations}
\subsubsection{$G'=U(1)\times SU(N)$}
Considering the case of $U(1)\times SU(N)$, the BPS-equations combined
with the Gauss law read
\begin{align}
\bar{\D}H &= 0 \ , \\
F_{12}^a t^a &= \frac{8\pi^2}{N\kappa\mu}
  \left(\Tr\left(H H^\dag\right) -\xi\right)
  \left(H H^\dag 
  -\frac{\mathbf{1}_N}{N}\Tr\left(H H^\dag\right)\right)
  \non&\phantom{=\ }
+\frac{8\pi^2}{\mu^2}
  \left[H H^\dag\left(H H^\dag 
  -\frac{\mathbf{1}_N}{N}\Tr\left(H H^\dag\right)\right)
  -\frac{\mathbf{1}_N}{N}\Tr\left(\left(H H^\dag\right)^2\right)
  +\frac{\mathbf{1}_N}{N^2}
  \left(\Tr\left(H H^\dag\right)\right)^2\right] \ , \nonumber\\
F_{12}^0 t^0 &= \frac{8\pi^2}{N^2\kappa^2}\Tr\left(H H^\dag\right)
  \left(\Tr\left(H H^\dag\right)-\xi\right)\mathbf{1}_N 
+\frac{8\pi^2}{N\kappa\mu}\left[
  \Tr\left(\left(H H^\dag\right)^2\right)
  -\frac{1}{N}\left(\Tr\left(H H^\dag\right)\right)^2\right]
  \mathbf{1}_N \ . \nonumber
\end{align}
In this case, the generic vacuum is given by 
\beq \langle H\rangle = \sqrt{\frac{\xi}{N}}\mathbf{1}_N \ . 
\label{eq:colorflavorvacuum} \eeq 
This vacuum allows for an unbroken global symmetry, the so-called
color-flavor symmetry which is the global part of the gauge
transformation combined with the flavor symmetry. This is of crucial
importance for having orientational modes in vortex configurations. 

Utilizing the moduli matrix formalism, we can immediately solve the
first BPS-equation and rewrite the second in terms of the new
variables
\begin{align} 
H &= S^{-1}H_0(z) \ , \quad 
\bar{A}^a t^a = - i {S'}^{-1}\bar{\p}S' \ , \quad
\bar{A}^0 t^0 = -i \bar{\p}\log s
\end{align}
along with the definitions $\Omega \equiv \omega\Omega', 
\Omega' \equiv S' {S'}^\dag, \omega \equiv s s^\dag$ and $\Omega_0
\equiv H_0(z)H_0^\dag(z)$. The field-strength matrices are 
\begin{align} 
F_{12}^a t^a = 
2{S'}^{-1}\bar{\p}\left[\Omega'\p{\Omega'}^{-1}\right]S' \ , 
\qquad
F_{12}^0 t^0 =
-2\mathbf{1}_N \bar{\p}\p\log\omega \ . \label{eq:f12}
\end{align}
In this $U(1)\times SU(N)$ case we can write down the two master
equations like
\begin{align}
\bar{\p}\left[\Omega'\p{\Omega'}^{-1}\right] &=
\frac{4\pi^2}{N\kappa\mu}
  \frac{1}{\omega}
  \left(\frac{1}{\omega}\Tr\left(\Omega_0{\Omega'}^{-1}\right) -\xi\right)
  \left(\Omega_0{\Omega'}^{-1}
  -\frac{\mathbf{1}_N}{N}\Tr\left(\Omega_0{\Omega'}^{-1}\right)\right)
  \non&\phantom{=\ }
+\frac{4\pi^2}{\mu^2}
  \frac{1}{\omega^2}\bigg[\Omega_0{\Omega'}^{-1}\left(\Omega_0{\Omega'}^{-1} 
  -\frac{\mathbf{1}_N}{N}\Tr\left(\Omega_0{\Omega'}^{-1}\right)\right)
  \non&\phantom{=\ \frac{4\pi^2}{\mu^2}\frac{1}{\omega^2}\bigg[\ }
  -\frac{\mathbf{1}_N}{N}\Tr\left(\left(\Omega_0{\Omega'}^{-1}\right)^2\right)
  +\frac{\mathbf{1}_N}{N^2}
  \left(\Tr\left(\Omega_0{\Omega'}^{-1}\right)\right)^2\bigg] \ , \\
\bar{\p}\p\log\omega &= -\frac{4\pi^2}{N^2\kappa^2}
  \frac{1}{\omega}\Tr\left(\Omega_0{\Omega'}^{-1}\right)
  \left(\frac{1}{\omega}\Tr\left(\Omega_0{\Omega'}^{-1}\right)-\xi\right) 
  \non&\phantom{=\ }
+\frac{4\pi^2}{N\kappa\mu}\frac{1}{\omega^2}\left[
  \Tr\left(\left(\Omega_0{\Omega'}^{-1}\right)^2\right)
  -\frac{1}{N}\left(\Tr\left(\Omega_0{\Omega'}^{-1}\right)\right)^2\right]
  \ . 
\end{align}
Setting the couplings equal $\kappa=\mu$, we can write the $U(N)$
Chern-Simons BPS-equations and master equation as simple as
\begin{align}
F_{12}^\alpha t^\alpha &= \frac{8\pi^2}{\kappa^2} H H^\dag
\left(H H^\dag - \frac{\xi}{N}\mathbf{1}_N\right) \ , \\
\bar{\p}\left[\Omega\p\Omega^{-1}\right] &=
\frac{4\pi^2}{\kappa^2}\Omega_0\Omega^{-1}
\left[\Omega_0\Omega^{-1} - \frac{\xi}{N}\mathbf{1}_N\right] \ .
\end{align}
The boundary conditions for these master equations coincide with the
weak coupling solutions (\ref{eq:weakSU}).

\subsubsection{$G'=U(1)\times SO(N)$ and $G'=U(1)\times USp(2M)$}
Considering now the gauge group $G=U(1)\times SO(N)$ and $G=U(1)\times
USp(2M)$ on the same footing with their corresponding invariant tensor
$J$, which has the properties $J^\dag J = \mathbf{1}_N$ and $J^{\rm T}
= \epsilon J$ with $\epsilon = \pm 1$ for $SO(N)$ and $USp(2M)$,
respectively. 

The vacuum has the generic form \cite{Eto:2008qw}
\beq \langle H\rangle = \diag\left(v_1, v_2, \ldots, v_N \right) \ , \quad v_i \in \mathbb{R}_+ \ , \eeq
however, we will consider the most symmetric vacuum allowing for the
global color-flavor symmetry, viz.~we will here use
(\ref{eq:colorflavorvacuum}). 
We have the following system which is obtained by
combining the BPS equations with the Gauss law and applying respective
algebras 
\begin{align}
\bar{\D}H &= 0 \ , \\
F_{12}^a t^a &= \frac{4\pi^2}{N\kappa\mu}
  \left(\Tr\left(H H^\dag\right) -\xi\right)
  \left(H H^\dag - J^\dag\left(H H^\dag\right)^{\rm T}J\right)
+\frac{2\pi^2}{\mu^2}
  \left[\left(H H^\dag\right)^2
  -J^\dag\left(\left(H H^\dag\right)^2\right)^{\rm T}J\right] \ , 
  \nonumber \\ 
F_{12}^0 t^0 &= \frac{8\pi^2}{N^2\kappa^2}\Tr\left(H H^\dag\right)
  \left(\Tr\left(H H^\dag\right)-\xi\right)\mathbf{1}_N 
+\frac{4\pi^2}{N\kappa\mu}
  \Tr\left(H H^\dag\left(H H^\dag 
  -J^\dag \left(H H^\dag\right)^{\rm T}J\right)\right)\mathbf{1}_N
  \ ,
  \nonumber
\end{align}
which lead to the master equations
\begin{align}
\bar{\p}\left[\Omega'\p{\Omega'}^{-1}\right] &= 
  \frac{2\pi^2}{N\kappa\mu}\frac{1}{\omega}
  \left(\frac{1}{\omega}\Tr\left(\Omega_0{\Omega'}^{-1}\right) -\xi\right)
  \left(\Omega_0{\Omega'}^{-1} 
  -J^\dag\left(\Omega_0{\Omega'}^{-1}\right)^{\rm T}J\right)
  \non&\phantom{=\ }
+\frac{\pi^2}{\mu^2}\frac{1}{\omega^2}
  \left[\left(\Omega_0{\Omega'}^{-1}\right)^2
  -J^\dag\left(\left(\Omega_0{\Omega'}^{-1}\right)^2\right)^{\rm T}J\right] \ , 
  \\ 
\bar{\p}\p\log\omega &= -\frac{4\pi^2}{N^2\kappa^2}
  \frac{1}{\omega}\Tr\left(\Omega_0{\Omega'}^{-1}\right)
  \left(\frac{1}{\omega}
  \Tr\left(\Omega_0{\Omega'}^{-1}\right)-\xi\right)
  \non&\phantom{=\ }
-\frac{2\pi^2}{N\kappa\mu}\frac{1}{\omega^2}
  \Tr\left(\Omega_0{\Omega'}^{-1}\left(\Omega_0{\Omega'}^{-1} 
  -J^\dag \left(\Omega_0{\Omega'}^{-1}\right)^{\rm T}J\right)\right)
  \ .
\end{align} 
The boundary conditions for these master equations coincide with the
weak coupling solutions (\ref{eq:weakSOUSp}). 

\subsubsection{Energy density and flux densities}

Rewriting the energy density (\ref{eq:energydensity1}) in terms of our
new variables and remembering the boundary term which vanishes when
integrating over the entire plane, while nevertheless produces a
big difference between the magnetic flux density and the energy
density, we have
\beq \mathcal{E} = 2\xi\bar{\p}\p\log\omega 
  + 2\bar{\p}\p\left(\frac{1}{\omega}\Tr\Omega_0{\Omega'}^{-1}\right)
  \ . \eeq 
However, the total energy
\beq E = \int_{\mathbb{C}} \mathcal{E} = 2\pi\xi\nu = \frac{2\pi\xi
  k}{n_0} \ , \eeq
is simply proportional to the topological charge as always.

The Abelian magnetic flux density is the first term (up to a factor)
in the energy density 
\beq \mathcal{B} = F_{12}^0 = -2\sqrt{2N}\;\bar{\p}\p\log\omega \ ,
\eeq
whereas the non-Abelian flux is the matrix defined in
Eq.~(\ref{eq:f12}).  
The Abelian electric field density reads
\beq E_i = F_{i0}^0 = \frac{2\pi}{\kappa}
\sqrt{\frac{2}{N}}\p_i
\left(\frac{1}{\omega}\Tr\left(\Omega_0{\Omega'}^{-1}\right)\right)
\ , \eeq
while the non-Abelian electric field density is given by
\beq E_i^a t^a = F_{i0}^a t^a = 
\frac{4\pi}{\mu}\p_i\Tr\left(H H^\dag t^a\right) t^a \ . \eeq 
This can be written for $G'=SU(N)$ as
\beq E_i^a t^a = \frac{2\pi}{\mu}\p_i\left[\frac{1}{\omega}\left(
{S'}^{-1}\Omega_0{\Omega'}^{-1}S' 
-\frac{1}{N}\Tr\left(\Omega_0{\Omega'}^{-1}\right)\right)\right] \ ,
\eeq
while for $G'=SO(N)$ or $G'=USp(2M)$ it is
\beq E_i^a t^a = \frac{\pi}{\mu}\p_i\left[\frac{1}{\omega}
{S'}^{-1}\left(\Omega_0{\Omega'}^{-1} 
-J^\dag\left(\Omega_0{\Omega'}^{-1}\right)^{\rm T}J\right)S'\right] \ .
\eeq

\subsection{Solutions}

In the Abelian Chern-Simons theory, there exists a rigorous existence
proof of the solutions in Ref.~\cite{Wang:1991na}. To our knowledge
this has not rigorously been proved in the theory at hand.
In the case of the vortices in the Yang-Mills-Higgs theory,
the ``covariant holomorphic'' condition on the Higgs fields
$\bar{D}H=0$, which is solved by the moduli matrix formalism,
does uniquely determine the full moduli space of vortices via the
Hitchin-Kobayashi correspondence 
\cite{Mundet i
  Riera:1999fd,cieliebak,Baptista:2004rk,Baptista:2008ex}, which
however has only been proved on compact spaces. 
This means that the corresponding master equations do not induce
further moduli. For the vortices with the $U(N)$ gauge group, an index
theorem has been given in Ref.~\cite{Hanany:2003hp} while for generic
gauge groups (under certain conditions) an index theorem has been
given in Ref.~\cite{Eto:2009bg}. The index computed gives the number
of moduli and does indeed correspond to the number of moduli found in
the moduli matrix. 

The first part of constructing a solution is to write down the moduli
matrix. Here we simply follow the way paved by
the paper \cite{Eto:2008yi} using holomorphic invariants of the gauge
subgroup $G'$. This boils down to some constraints for the moduli
matrix to obey. A few examples of interest here is the case of
$G'=SU(N)$
\beq \det H_0(z) = z^k + \mathcal{O}\left(z^{k-1}\right) \ , 
\label{eq:SUconstraint} \eeq
while in the case of $G'=SO,USp$, respectively, we have
\beq H_0^{\rm T}(z) J H_0(z) = z^{\frac{2k}{n_0}} J +
\mathcal{O}\left(z^{\frac{2k}{n_0}-1}\right) \ , 
\label{eq:SOUSpconstraint} \eeq
where $k$ is the vortex number (recall that $\nu = \frac{k}{n_0}$ is
the $U(1)$ winding) and $n_0=2$ in case of $SO(2M)$ and $USp(2M)$
while $n_0=1$ for $SO(2M+1)$, $M$ being positive integers. For
$SU(N)$, however $n_0=N$.

The rather complicated
looking master equations found in the last section are assumed to have  
a unique solution for each moduli matrix $H_0(z)$ (up to $V$
equivalence, see Ref.~\cite{Eto:2008yi}). That is the moduli matrices
are redundant and have to be identified by the following $V$
transformation 
\beq H_0(z) \sim V(z,\bar{z}) H_0(z) \ , \quad
S(z,\bar{z}) \sim V(z,\bar{z})S(z,\bar{z}) \ , \quad
V \in G^{\mathbb{C}} \ .  \eeq
Here we conjecture the existence and uniqueness of the solutions to
the master equation for each moduli matrix (up to the $V$
equivalence). To provide plausibility for this claim we shall continue
in two directions. 

First we consider the weak coupling limit $\kappa\to 0$ and
$\mu\to 0$, which seems like an odd limit to take, but having an
advantage. Looking at the theory (\ref{eq:LYMCSH}) it is immediately
seen 
that the matter fields are forced to stay in the vacuum manifold
corresponding to the \emph{strong} coupling limit of the normal
non-Abelian vortex (i.e.~with only a Yang-Mills kinetic term). In
turn, this gives us a unique solution which in fact is the same
solution as found in the strong coupling limit of the non-Abelian
vortex with only a Yang-Mills kinetic term. This solution, appropriate
only for vortices of the semi-local type, are usually called lumps in
the literature. 

The second direction we will take will simply be to find some
solutions by numerical calculations. 

Now the existence of the solutions to the master equations, as we
argue, makes it possible to exploit a lot of results developed in the 
literature. In short,
\begin{align} \begin{array}{c}
\textrm{\it the moduli space of non-Abelian Chern-Simons $k$ vortices
  with gauge group $G$}\\\textrm{\it
  is equal to the moduli space of the non-Abelian Yang-Mills
  $k$ vortices with gauge group $G$.} \end{array} 
\end{align}
Moduli spaces of the non-Abelian vortices in ${\cal N}=2$ sQCD has
been found in the literature in Refs.~\cite{Hanany:2003hp,Eto:2005yh}
for $U(N)$ and in Refs.~\cite{Eto:2009bg} for $SO(N),USp(2M)$. 

Here we will summarize a few results from the literature. In the
pioneering papers \cite{Hanany:2003hp,Auzzi:2003fs} discovering the
non-Abelian vortices with gauge group $U(N)$ (in contrast to the
formerly found $\mathbb{Z}_N$ strings) the moduli space of a single
vortex string was found to be \beq \mathcal{M}_{k=1,G'=SU(N)} =
\mathbb{C}\times \mathbb{C}P^{N-1} , \eeq 
where the first factor denotes the position in the transverse plane
while the second factor are orientational modes. For well separated
$k$ vortices, the moduli space can be composed as simply the symmetric
product of that of the single vortex. This is not the case, when the
centers coincide. In the $k=2$, $U(2)$ case, the moduli space has been
found explicitly in the Refs.~\cite{Auzzi:2005gr,Eto:2006cx}
\beq \mathcal{M}_{k=2,G'=U(2)} = \mathbb{C}\times
W\mathbb{C}P^2_{2,1,1} \ ,  \eeq which decomposes into a
center-of-mass position and a weighted complex projective space with
unequal weights giving rise to a conical type of singularity. 
In Ref.~\cite{Eto:2009bg} the moduli spaces of vortices with gauge
groups $G=U(1)\times SO(N)$ and $G=U(1)\times USp(2M)$ has been
found. A complication arises due to the fact that already for $\NF=N$
flavors, the vortices are in general of the semi-local type (i.e.~they
have polynomial tails in their profile functions). The spaces
quoted here correspond to the vortices of local type, thus they are
constrained to have holomorphic invariants with coincident zeroes. In
the language of Ref.~\cite{Eto:2009bg} this is obtained by
constraining the vortices by the so-called strong condition
\beq H_0^{\rm T}(z) J H_0(z) = \left(z-z_0\right)^{\frac{2k}{n_0}} J
\ . \eeq 
The single local vortex with $G'=USp(2M)$ has the moduli space
\beq \mathcal{M}_{k=1,G'=USp(2M)} = \mathbb{C}\times \frac{USp(2M)}{U(M)}
\ , \eeq
while in the case of $G'=SO(2M)$ it is found to be
\beq \mathcal{M}_{k=1,G'=SO(2M)} = \left(\mathbb{C}\times
\frac{SO(2M)}{U(M)}\right)_+ \cup \left(\mathbb{C}\times
\frac{SO(2M)}{U(M)}\right)_- \ , \eeq
where the $\pm$ denotes the chirality as described in detail in
Ref.~\cite{Eto:2009bg} which is deeply rooted in the fact that the
first homotopy group has in addition to the integers a $\mathbb{Z}_2$
factor. This can also be interpreted as two spinor representations
which is exactly the irreducible representations of the dual group
$\tilde{G'}$, where the dual is defined as being the group having the
root vectors 
$\vec\alpha^*= \frac{\vec\alpha}{\vec\alpha\cdot\vec\alpha}$. 
For the $k=2$, $G'=SO(2M)$ the following orientational moduli spaces
have been found to be locally
\begin{align}
\mathcal{M}_{k=2,G'=SO(4m),Q_{\mathbb{Z}_2}=+1} &= \mathbb{R}_+^m\times
\frac{SO(4m)}{USp(2)^m} \times \mathbb{Z}_2 \ , \\
\mathcal{M}_{k=2,G'=SO(4m),Q_{\mathbb{Z}_2}=-1} &= \mathbb{R}_+^{m-1}\times
\frac{SO(4m)}{U(1)\times USp(2)^{m-1}\times SO(2)} \ , \\
\mathcal{M}_{k=2,G'=SO(4m+2),Q_{\mathbb{Z}_2}=+1} &= \mathbb{R}_+^m\times
\frac{SO(4m+2)}{U(1)\times USp(2)^m} \times \mathbb{Z}_2 \ , \\
\mathcal{M}_{k=2,G'=SO(4m+2),Q_{\mathbb{Z}_2}=-1} &= \mathbb{R}_+^m\times
\frac{SO(4m+2)}{USp(2)^m\times SO(2)} \ .
\end{align}
In the case of $k=1$, $G'=SO(2M+1)$, the moduli spaces are quite
similar to the $k=2$ even case.

\subsection{Weak coupling limit\label{sec:weakcoup}}
\subsubsection{$G'=U(1)\times SU(N)$}
Taking $\kappa=\mu\to 0$, we obtain from the $D$ term conditions 
\beq \Omega'=\left(\det\Omega_0\right)^{-\frac{1}{N}}\Omega_0 \ , \quad
\omega = \frac{N}{\xi}\left(\det\Omega_0\right)^{\frac{1}{N}} \ ,
\quad \Omega = \frac{N}{\xi}\Omega_0 \ , \label{eq:weakSU}\eeq
which can be packaged together as a $U(N)$ field $\Omega$. Instead of
taking both couplings simultaneously to weak coupling, we can play a
game of taking only one of them, keeping the other finite
(non-infinitesimal). Taking $\kappa\to 0$ and keeping $\mu$ finite we
obtain
\beq \omega = \frac{1}{\xi}\Tr\Omega_0{\Omega'}^{-1} \ , \eeq
at the zeroth order in $\kappa$ while at first order we get the
constraint
\beq N\Tr\left(\left(\Omega_0{\Omega'}^{-1}\right)^2\right) = 
\left(\Tr\,\Omega_0{\Omega'}^{-1}\right)^2 \ . \eeq
We note that only the Abelian field is determined, however at first
order in the coupling constant we obtain a single constraint on the
non-Abelian fields.
Taking instead $\mu\to 0$ keeping $\kappa$ finite we have
\beq \Omega' = \Lambda\Omega_0 \ , \qquad {\rm with}\ \Lambda \in {\rm
  const.} \ , \eeq
to both zeroth and first order in $\mu$. 

\subsubsection{$G'=U(1)\times SO(N)$ and $G'=U(1)\times USp(2M)$}
Taking $\kappa=\mu\to0$ we have from the $D$ term conditions
\cite{Eto:2009bg,Eto:2008qw} 
\beq \Omega' = H_0(z)\frac{\mathbf{1}_N}{\sqrt{M^\dag M}}H_0^\dag(z)
\ , \qquad \omega = \frac{1}{\xi}\Tr\sqrt{M^\dag M} \ ,  
\label{eq:weakSOUSp}\eeq
where $M = H_0^{\rm T}(z) J H_0(z)$ is the meson field of the $SO,USp$
theories according to the choice of the gauge group and in turn
invariant tensor. 

A comment in store is that the Chern-Simons term is simply switched
off in this limit and the lumps are \emph{the same} as the ones living
in the Yang-Mills theories experiencing infinitely massive gauge
bosons. The point here, however, is to argue by continuity the
existence and uniqueness of the solutions to the master equations for
a given moduli matrix $H_0(z)$ (up to the $V$-equivalence relation).

\subsection{Numerical solutions}

\subsubsection{Example: $U(N)$}
Let us do a warm-up and consider the single $U(N)$ Chern-Simons vortex
($\kappa=\mu$) as has been found in
Refs.~\cite{Lozano:2007yz,Aldrovandi:2007nb}, however doing it in our
formalism.  
Taking a simple moduli matrix
\beq H_0(z) = \diag\left(z,\mathbf{1}_{N-1}\right) \ , \eeq
which of course satisfies the constraint (\ref{eq:SUconstraint}), thus
we can use the Ansatz for $\Omega$
\beq \Omega = e^{\psi}
\diag\left(e^{(N-1)\chi},e^{-\chi}\mathbf{1}_{N-1}\right) \ , \eeq
leading to the two coupled equations of motion
\begin{align}
\bar{\p}\p\left[\psi+(N-1)\chi\right] &= 
-\frac{4\pi^2}{\kappa^2}\left|z-z_0\right|^2e^{-\psi-(N-1)\chi}
\left(\left|z-z_0\right|^2 e^{-\psi-(N-1)\chi}-\frac{\xi}{N}\right) \ , \\
\bar{\p}\p\left[\psi-\chi\right] &= - \frac{4\pi^2}{\kappa^2}
e^{-\psi+\chi}\left(e^{-\psi+\chi} - \frac{\xi}{N}\right) \ .
\end{align}
Notice that the two equations decouple in the sense that there only
appear the combinations $\psi+(N-1)\chi$ and $\psi-\chi$. In fact it
is easily seen that in this case, the field combination $\psi-\chi$
can be in the vacuum in all $\mathbb{C}$ which trivially solves the
second equation. However, the first equation still needs to be solved
numerically. 
The boundary conditions are
\beq \psi_{\infty} = \log\left(\frac{N|z|^{\frac{2}{N}}}{\xi}\right)
\ , \qquad 
\chi_{\infty} = \log\left(|z|^{\frac{2}{N}}\right)\ . \eeq
The equations become essentially Abelian when the couplings are equal
$\kappa=\mu$, as was noted in Ref.~\cite{Lozano:2007yz}. 
The energy density is given by
\beq \mathcal{E} = 2\xi\bar{\p}\p\psi
+2\bar{\p}\p\left[|z|^2 e^{-\psi-(N-1)\chi} +
(N-1)e^{-\psi+\chi}\right] \ , \eeq
where the last term is the boundary term which of course integrates to
zero. 
The Abelian and non-Abelian magnetic flux densities are given by 
\beq F_{12}^0 = -2\sqrt{2N}\;\bar{\p}\p\psi \ , \quad
F_{12}^a t^a = -2\sqrt{2N(N-1)}\;\bar{\p}\p\chi \, t \ , \eeq
where the following matrix has been defined for convenience 
\beq t \equiv \frac{1}{\sqrt{2N(N-1)}}
\diag\left(N-1,-\mathbf{1}_{N-1}\right) \ , \eeq
which is traceless and has the trace of its square normalized to one
half.
The Abelian electric field density is 
\beq E_r = \frac{2\pi}{\kappa}\sqrt{\frac{2}{N}}
\p_r\left[r^2 e^{-\psi-(N-1)\chi} + (N-1)e^{-\psi+\chi}\right] \ ,
\eeq 
while the non-Abelian electric field density is
\beq E_r^a t^a = \frac{2\pi}{\mu N}\sqrt{2N(N-1)}\p_r
\left[r^2 e^{-\psi-(N-1)\chi} - e^{-\psi+\chi}\right] t \ . \eeq
We will find in the next subsection, that the numerical solution for this 
vortex for $N=2$ is up to rescaling of some parameters equivalent to
the vortex studied in the next subsection (when $\kappa=\mu$). Thus
the concrete graphs are shown only for the vortex solution below.

\subsubsection{Example: $U(1)\times SO(2M)$ and $U(1)\times USp(2M)$} 

\begin{figure}[!tbp]
\begin{center}
\mbox{
\subfigure[]{\resizebox{!}{5.2cm}{\includegraphics{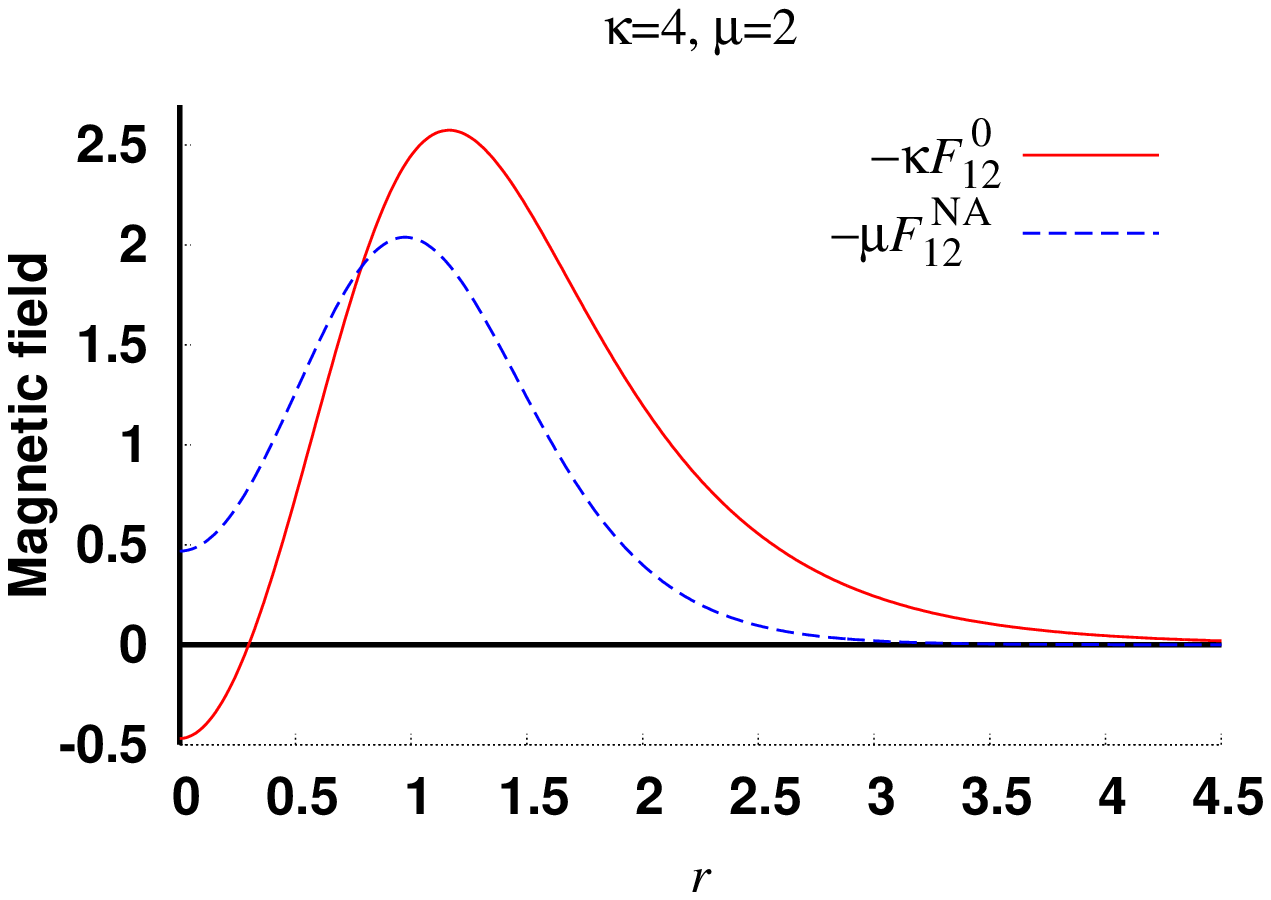}}}\quad
\subfigure[]{\resizebox{!}{5.2cm}{\includegraphics{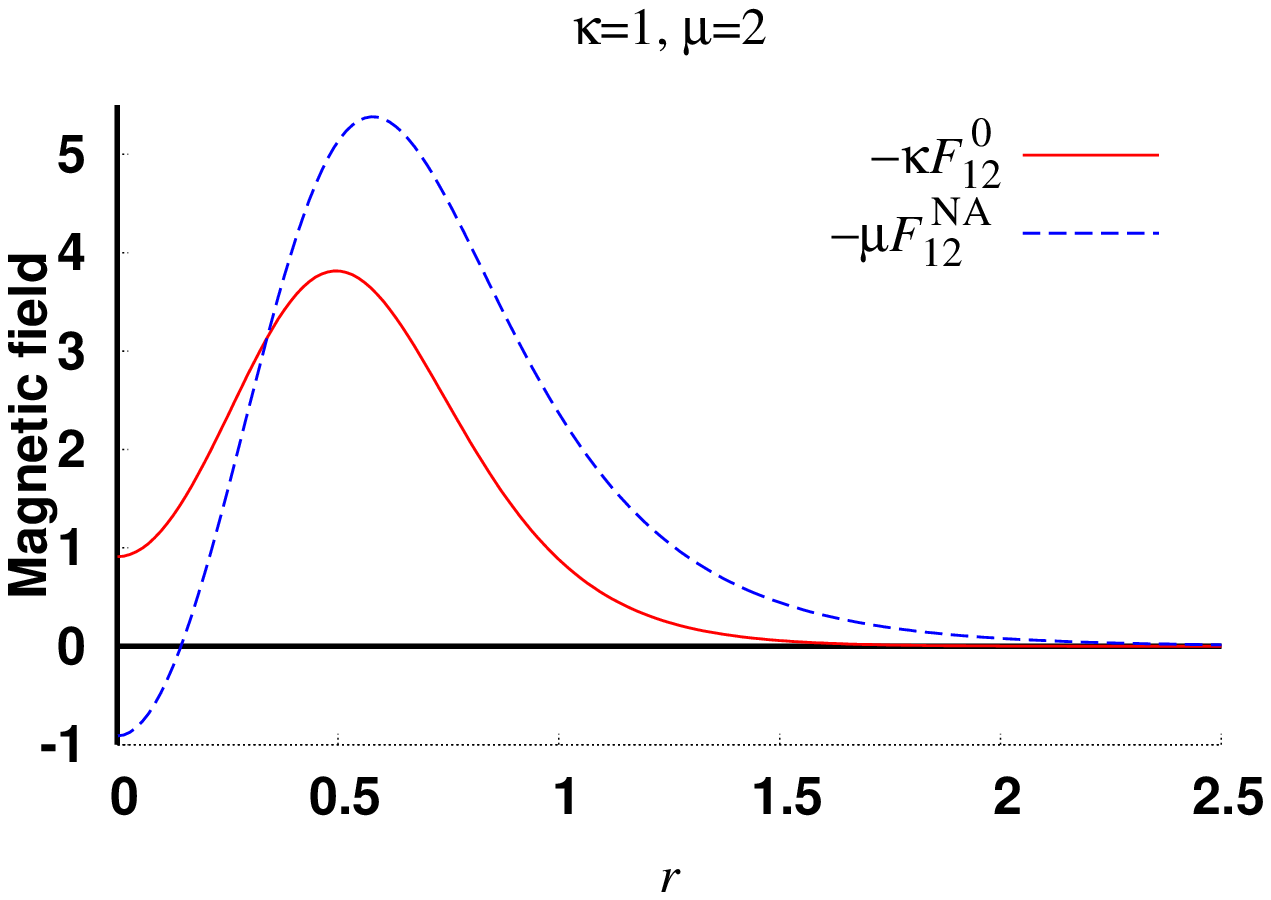}}}}
\caption{Differently normalized Abelian and non-Abelian magnetic fields
as $\kappa F_{12}^0$ and $\mu F_{12}^{\rm NA}$ for (a) $\kappa=4,\mu=2$ and
(b) $\kappa=1,\mu=2$. This combination cancels exactly at the origin
(to a numerical accuracy better than $\sim 10^{-5}$). The FI parameter
$\xi=2$. } 
\label{fig:type12}
\end{center}
\end{figure}

Let us take a simple example of a moduli matrix
\beq H_0(z) = \diag\left(z\mathbf{1}_M,\mathbf{1}_M\right) \ , \eeq
which surely satisfies the constraint (\ref{eq:SOUSpconstraint}). We
take the Ansatz
\beq \Omega' = \diag\left(e^\chi\mathbf{1}_M,
e^{-\chi}\mathbf{1}_M\right) \ , \quad \omega = e^\psi \ , \eeq 
where $\det\Omega'=1$ is manifest. The equations of motion in terms of
the new fields are
\begin{align}
\bar{\p}\p\chi =&\; - \frac{\pi^2}{\kappa\mu}
\left(|z|^2 e^{-\psi-\chi} + e^{-\psi+\chi} - \frac{\xi}{M}\right)
\left(|z|^2 e^{-\psi-\chi} - e^{-\psi+\chi}\right) \nonumber\\
&\;-\frac{\pi^2}{\mu^2}
\left(\left(|z|^2 e^{-\psi-\chi}\right)^2 -
\left(e^{-\psi+\chi}\right)^2\right) \ , 
\label{eq:SOUSpchiEOM}\\
\bar{\p}\p\psi =&\; - \frac{\pi^2}{\kappa^2}
\left(|z|^2 e^{-\psi-\chi} + e^{-\psi+\chi}\right)
\left(|z|^2 e^{-\psi-\chi} + e^{-\psi+\chi} - \frac{\xi}{M}\right)
\nonumber \\ 
&\;-\frac{\pi^2}{\kappa\mu}
\left(|z|^2 e^{-\psi-\chi} - e^{-\psi+\chi}\right)^2 \ .
\label{eq:SOUSppsiEOM}
\end{align}
It is interesting to note that under rescaling of the FI parameter
$\xi\to M\xi$, the above equations of motion are exactly the ones of
the $U(1)\times SU(2)$ theory with the Ansatz used in the last
section.  
The boundary conditions are
\beq \psi_\infty = \log\left(\frac{2M}{\xi}|z|\right) \ , \quad 
\chi_\infty = \log\left(|z|\right) \ , \eeq
and the energy density reads
\beq \mathcal{E} = 2\xi\bar{\p}{\p}\psi 
+2M\bar{\p}\p\left[|z|^2e^{-\psi-\chi}+e^{-\psi+\chi}\right] \ , 
\label{eq:SOUSpenergydensity}\eeq
where the first term is proportional to the Abelian magnetic flux
density 
\beq F_{12}^0 = -4\sqrt{M}\;\bar{\p}\p\psi \ , \eeq
and the last is the boundary term which integrates to
zero, while the non-Abelian magnetic field density reads
\beq F_{12}^a t^a \equiv F_{12}^{\rm NA} t = 
-4\sqrt{M}\;\bar{\p}\p\chi \, t \ , \quad 
t \equiv\frac{1}{2\sqrt{M}}
\diag\left(\mathbf{1}_M,-\mathbf{1}_M\right) \ . 
\label{eq:SOUSp4namagflux}\eeq
The Abelian electric field density reads
\beq E_r = \frac{2\pi\sqrt{M}}{\kappa}
\p_r\left[r^2e^{-\psi-\chi}+e^{-\psi+\chi}\right] \ ,
\eeq 
whereas the non-Abelian electric field density is
\beq E_r^a t^a \equiv E_{r}^{\rm NA} t = \frac{2\pi\sqrt{M}}{\mu}\p_r
\left[r^2 e^{-\psi-\chi} - e^{-\psi+\chi}\right] t \ .
\label{eq:SOUSp4naelecflux}\eeq
We show the vortex with this Ansatz corresponding to different values
of the coupling constants $\kappa,\mu$ in the following figures. Here
we will take for definiteness the group $G'$ to be $SO(4)$ or
$USp(4)$ hence $M=2$, which within the chosen Ansatz are
equivalent. We furthermore set $\xi=2$. The total energy is thus
(recall the Ansatz is for a single $k=1$ vortex)
\beq E = \int_{\mathbb{C}} \mathcal{E} = \pi\xi \ . \eeq
\begin{figure}[!tbp]
\begin{center}
\mbox{
\subfigure[]{\resizebox{!}{5.2cm}{\includegraphics{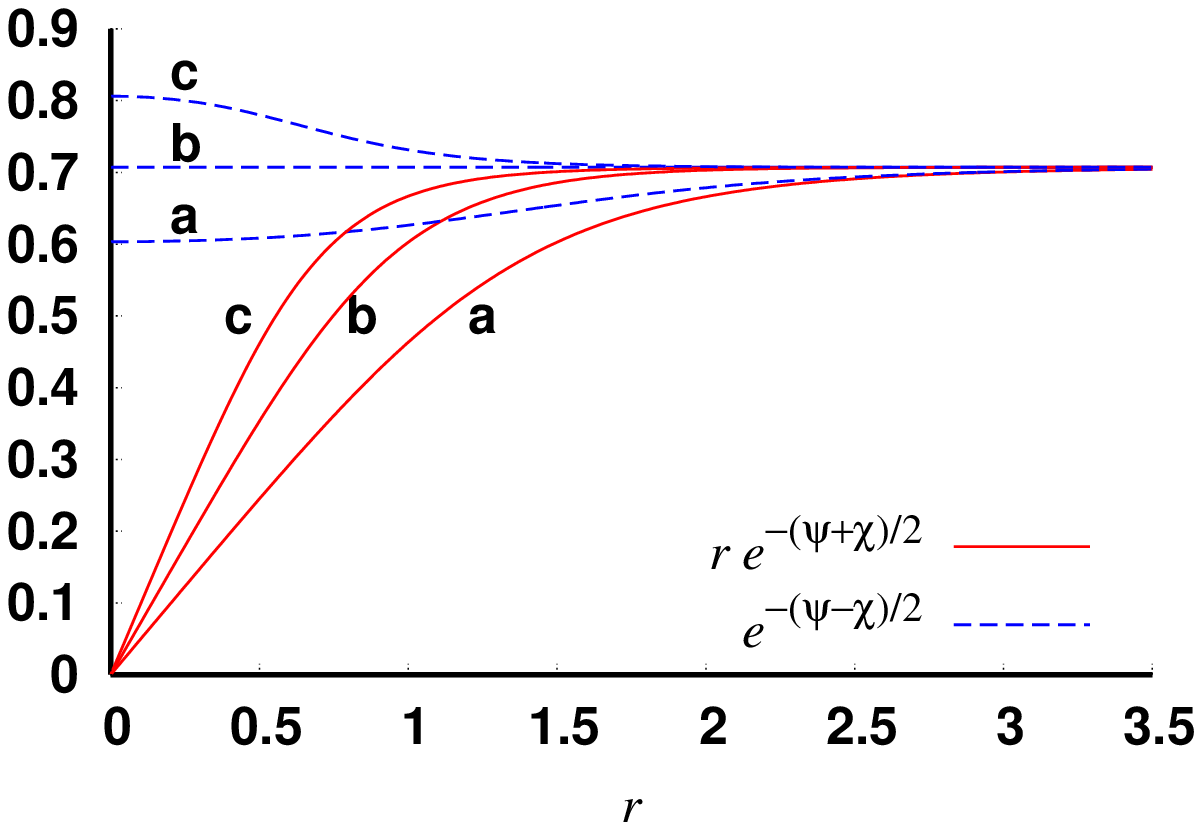}}}\quad
\subfigure[]{\resizebox{!}{5.2cm}{\includegraphics{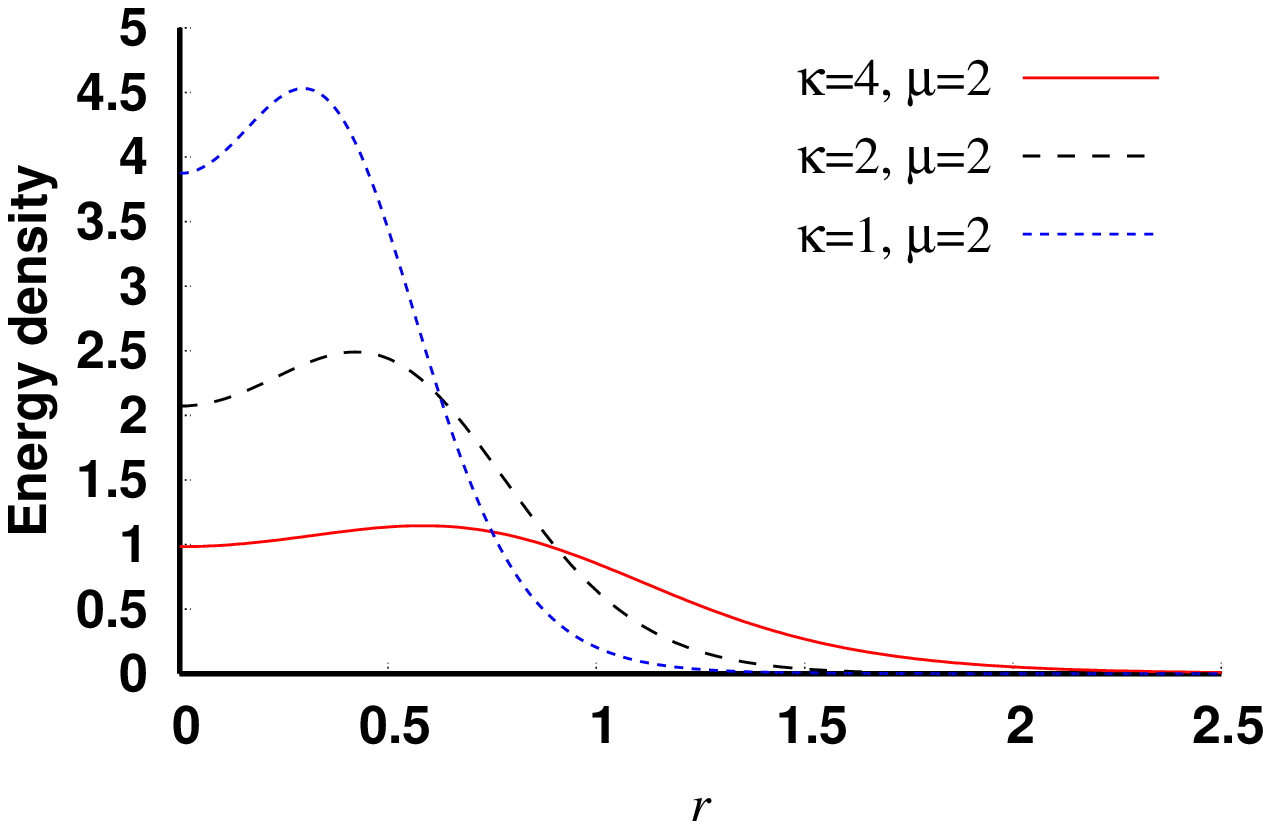}}}}
\caption{(a) Profile functions for three different values of the
  coupling constants; a: $\kappa=4,\mu=2$; b: $\kappa=2,\mu=2$; c:
  $\kappa=1,\mu=2$; the functions are plotted in traditional style
  with the winding field rising linearly and the non-winding field
  being constant at the origin. The FI parameter $\xi=2$. Notice
  that the VEV for these functions is $2^{-\frac{1}{2}}$. (b) The
  energy density $\mathcal{E}$ for the vortex for the same three
  different values of the couplings. All the energy densities
  integrate to 
  $\pi\xi$, within an accuracy better than $\sim 10^{-4}$. }
\label{fig:profiles+en}
\end{center}
\end{figure}
In Fig.~\ref{fig:profiles+en}a we show the profile
functions of the vortex in the traditional way, where the color-flavor
matrix is parametrized as follows
\beq H = \diag\left(f(r)e^{i\theta}\mathbf{1}_2,g(r)\mathbf{1}_2\right)
\ , \eeq 
which of course is equivalent to the parametrization in terms of
$\psi,\chi$. In Fig.~\ref{fig:profiles+en}b the energy density of
Eq.~(\ref{eq:SOUSpenergydensity}) is shown. The integral of the energy
density is identically equal to the integral of the Abelian magnetic
flux, as it should be. We see the vortex size is proportional to the
coupling constants. 
\begin{figure}[!tbp]
\begin{center}
\mbox{
\subfigure[]{\resizebox{!}{5.2cm}{\includegraphics{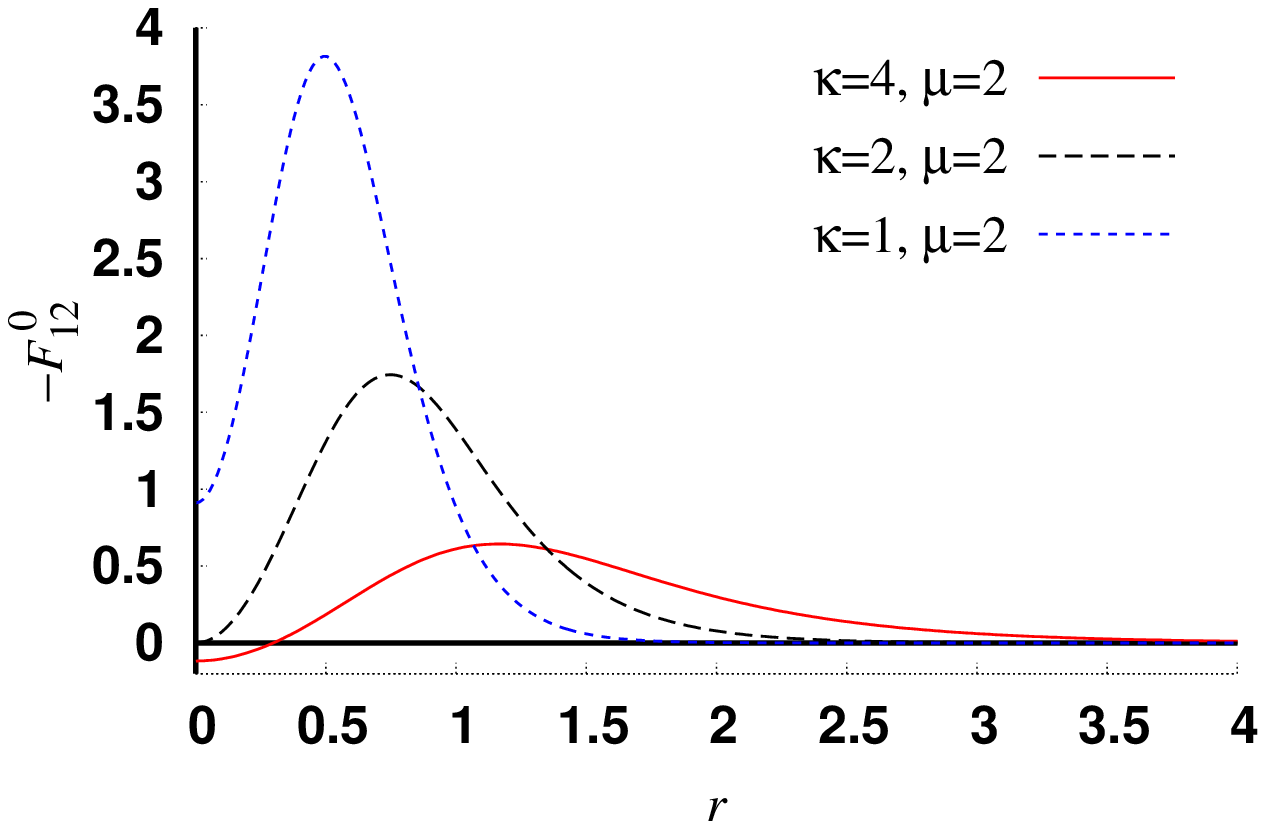}}}\quad
\subfigure[]{\resizebox{!}{5.2cm}{\includegraphics{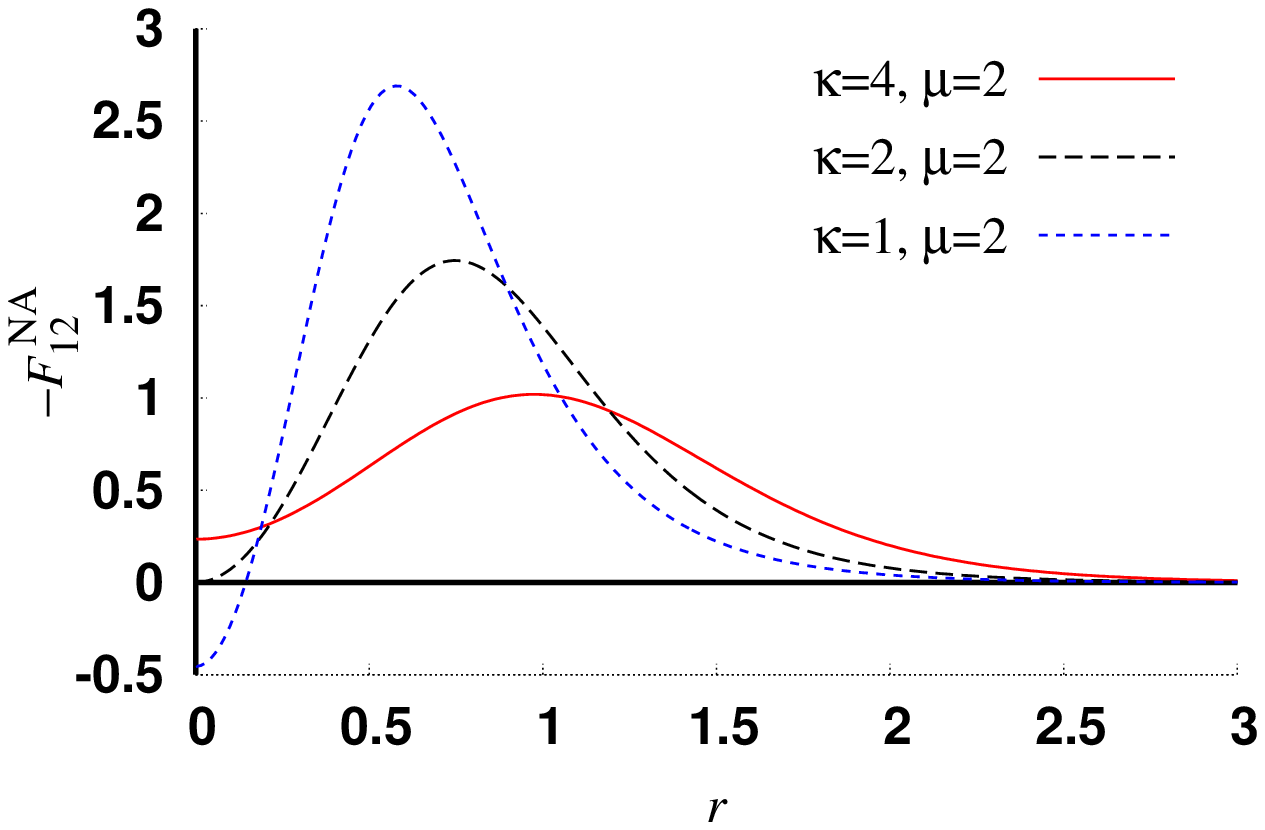}}}}
\caption{(a) The Abelian magnetic field $F_{12}^0$ (trace-part) for three
  different values of the couplings. Notice the equal coupling case
  has zero magnetic field at the origin while the different coupling
  cases have negative and positive values, respectively. (b) The
  non-Abelian magnetic field $F_{12}^a$ (traceless part) for different values
  of the couplings. Notice the opposite behavior of the
  non-Abelian magnetic field with respect the Abelian one at the
  origin, see also Fig.~\ref{fig:type12}. The FI parameter $\xi=2$. }
\label{fig:magflux}
\end{center}
\end{figure}
In Fig.~\ref{fig:magflux} we show the Abelian (a) and the non-Abelian
(b) magnetic field, respectively. We observe that the Abelian magnetic
field is negative at the origin while the non-Abelian magnetic field
is positive, in the $\kappa=4,\mu=2$ case. The contrary holds in the
$\kappa=1,\mu=2$ case where the non-Abelian magnetic field is negative
at the origin while the Abelian field is positive. 
It turns out that the combination 
\beq \left.\left(\kappa F_{12}^0 + 
\mu F_{12}^{\rm NA}\right)\right|_{r\to 0} = 0 \ .
\label{eq:normmagflux} \eeq 
An immediate consequence is that for
$|\kappa|\gg|\mu|$, $|F_{12}^{\rm NA}|\gg |F_{12}^0|$ at the origin
and vice versa.
Plots of the Abelian and non-Abelian magnetic fields normalized as
in Eq.~(\ref{eq:normmagflux}) are shown in Fig.~\ref{fig:type12} with
$\kappa=4,\mu=2$ in (a) and $\kappa=1,\mu=2$ in (b),
respectively. At the origin this combination cancels to a numerical
accuracy better than $10^{-5}$.  
\begin{figure}[!tbp]
\begin{center}
\mbox{
\subfigure[]{\resizebox{!}{5.2cm}{\includegraphics{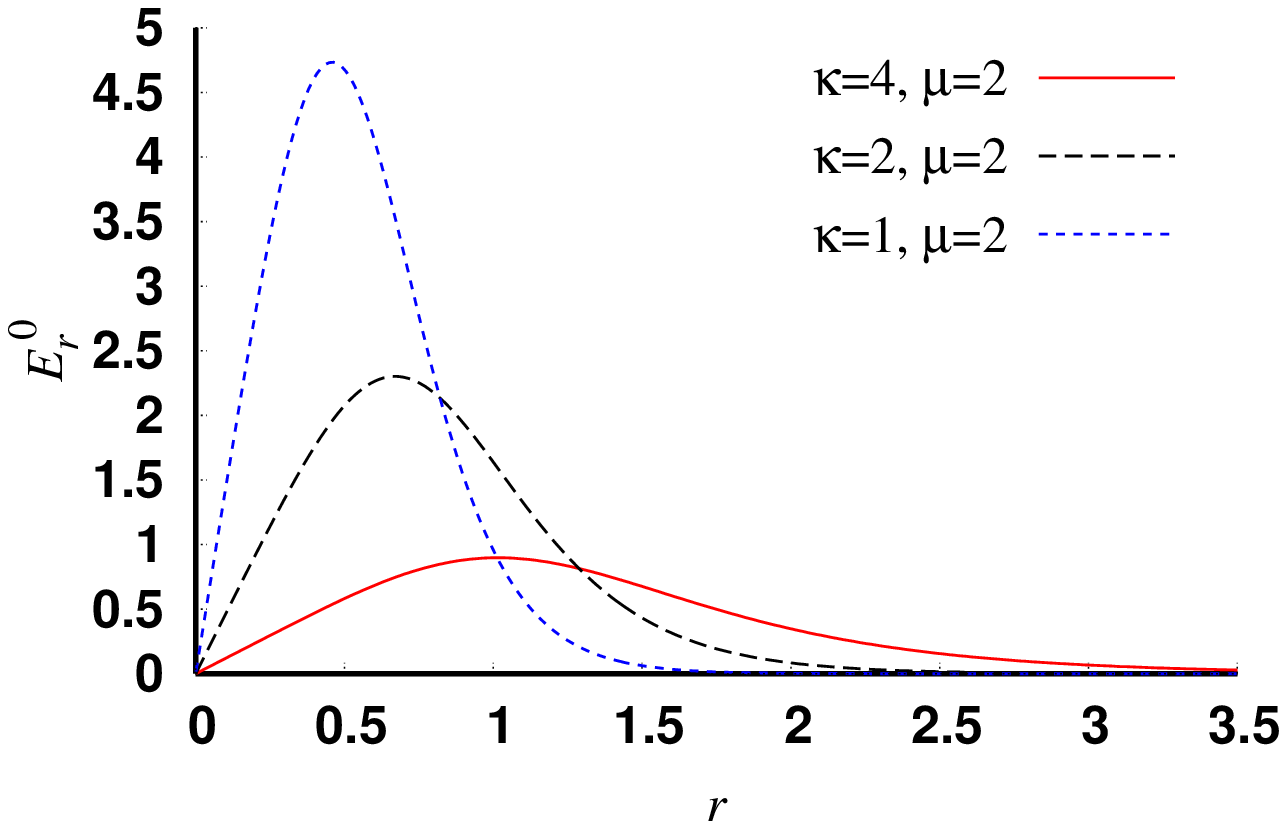}}}\quad
\subfigure[]{\resizebox{!}{5.2cm}{\includegraphics{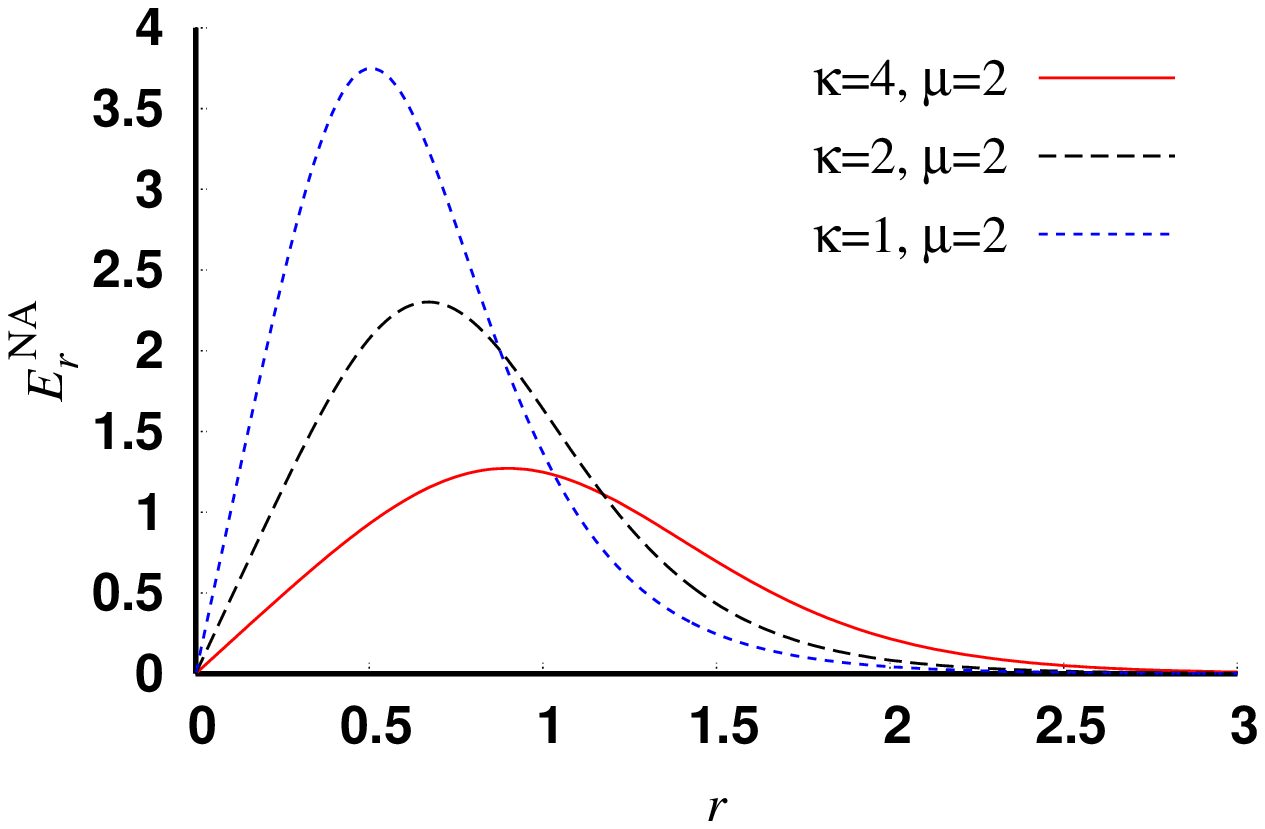}}}}
\caption{(a) The Abelian electric field in the radial direction
  $E_{r}$ (trace-part) for three different values of the
  couplings. (b) The non-Abelian electric field in the radial
  direction $E_{r}^{\rm NA}$ (traceless part). The FI parameter $\xi=2$. }
\label{fig:elecflux}
\end{center}
\end{figure}
First let us demonstrate the formula (\ref{eq:normmagflux}) by
calculating the fields in the limit $r\to 0$
\beq \left.\kappa F_{12}^0\right|_{r\to 0} =
\left.-\mu F_{12}^{\rm NA}\right|_{r\to 0} = 4\pi\sqrt{M}
\left[\frac{1}{\kappa}e^{-\psi+\chi}
\left(e^{-\psi+\chi}-\frac{\xi}{M}\right)
+\frac{1}{\mu}\left(e^{-\psi+\chi}\right)^2\right] \ . 
\label{eq:magfluxsplitting} \eeq
Note that the value of the magnetic fields only depends on the field
combination $\psi-\chi$, and it is understood that it has to be 
evaluated at the origin in the above equation. 
Secondly, let us demonstrate that the magnetic fields are zero at the
origin in the case of equal couplings. Subtracting
Eq.~(\ref{eq:SOUSpchiEOM}) from Eq.~(\ref{eq:SOUSppsiEOM}) we have
\begin{align} 
\bar{\p}\p\left(\psi-\chi\right) =
-\frac{\pi^2}{\kappa^2}\bigg[&
\left(1-\frac{\kappa^2}{\mu^2}\right)
\left(|z|^2 e^{-\psi-\chi}\right)^2
+\left(1-\frac{\kappa}{\mu}\right)
\left(2e^{-\psi+\chi}-\frac{\xi}{M}\right)|z|^2e^{-\psi-\chi} \nonumber\\
&+\left(1+\frac{\kappa}{\mu}\right)^2
\left(e^{-\psi+\chi}\right)^2
-\frac{\xi}{M}\left(1+\frac{\kappa}{\mu}\right)e^{-\psi+\chi} \bigg]
\ , \label{eq:subtracted}
\end{align}
which depends on $z,\bar{z}$ when the coupling constants are
different, $\kappa\neq\mu$. However, when the coupling constants are
equal, Eq.~(\ref{eq:subtracted}) reads
\beq \bar{\p}\p\left(\psi-\chi\right) =
-\frac{4\pi^2}{\kappa^2}
\left(e^{-\psi+\chi} - \frac{\xi}{2M}\right)e^{-\psi+\chi} \ , \eeq
which allows the field combination $\psi-\chi$ to stay constant with
the value 
\beq \psi-\chi = \log\left(\frac{2M}{\xi}\right) \ . \eeq
Plugging this (constant) solution into Eq.~(\ref{eq:magfluxsplitting})
we obtain readily $F_{12}^0=F_{12}^{\rm NA}=0$ in the limit $r\to 0$.

In Fig.~\ref{fig:elecflux} is shown the Abelian (a) and non-Abelian
(b) electric fields with different values of the couplings. 

\begin{figure}[!tbp]
\begin{center}
\mbox{
\subfigure[$\kappa>\mu$]{\resizebox{!}{2.5cm}{\includegraphics{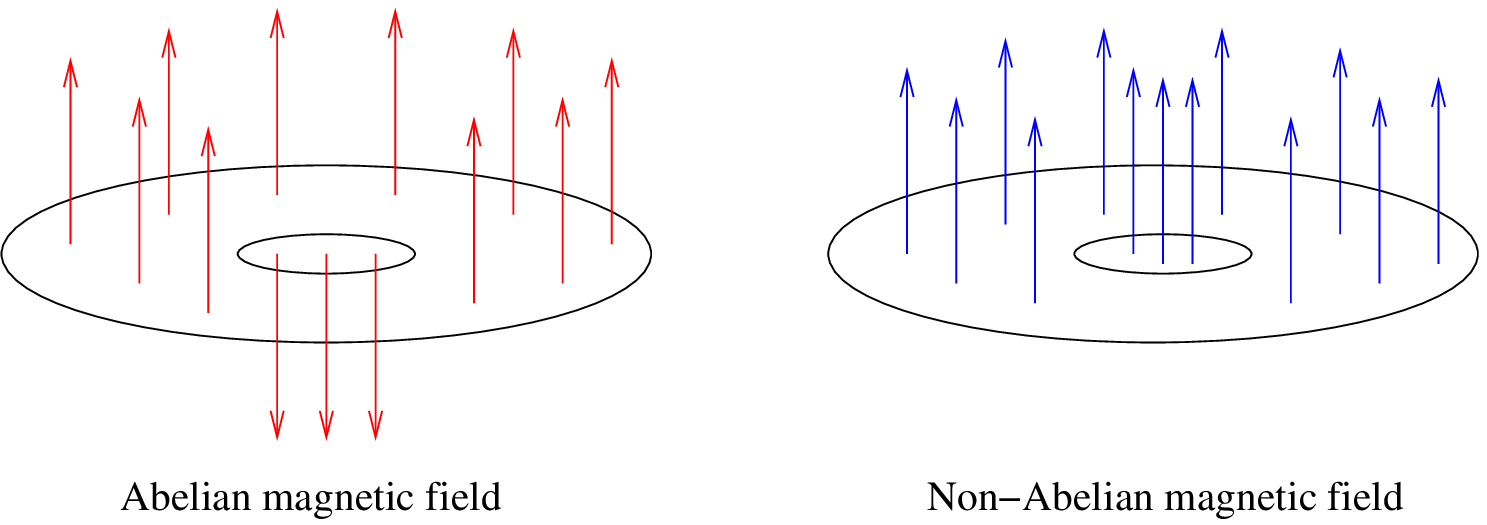}}}\quad
\subfigure[$\kappa<\mu$]{\resizebox{!}{2.5cm}{\includegraphics{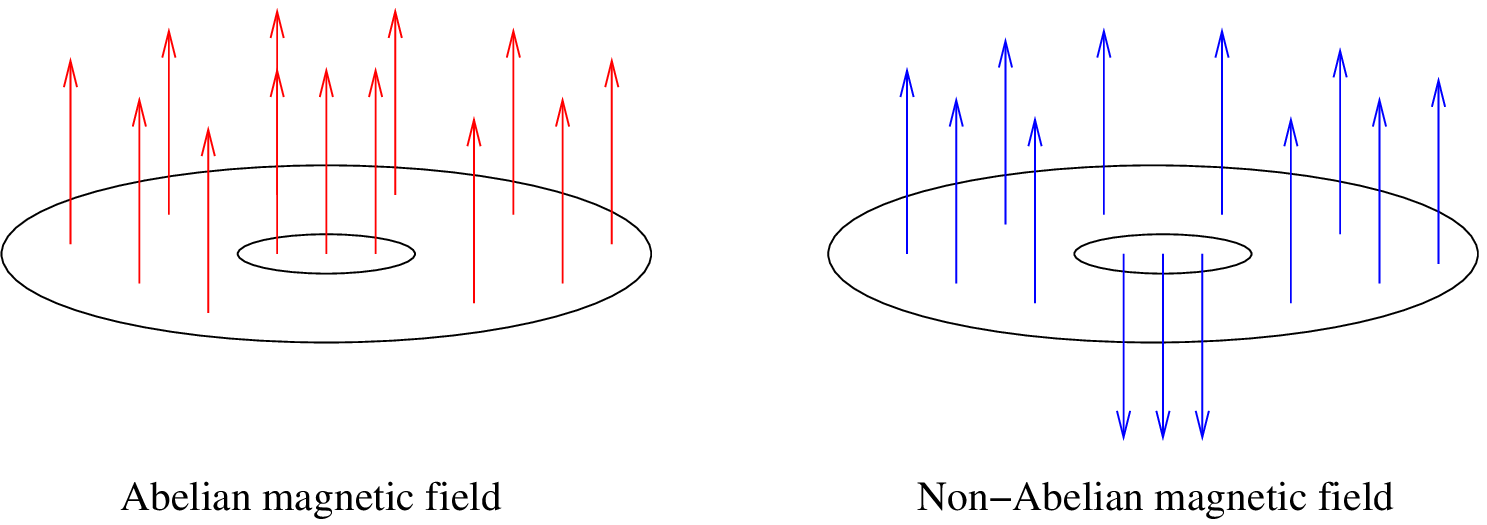}}}}
\caption{(a) Sketch of the magnetic fields where the Abelian
  (red/left) is negative at the origin and the non-Abelian
  (blue/right) is positive for $\kappa>\mu$. (b) Contrarily the
  Abelian (red/left) is positive at the origin while the non-Abelian
  (blue/right) is negative for $\kappa<\mu$. }
\label{fig:sketch_negative}
\end{center}
\end{figure}
In Fig.~\ref{fig:sketch_negative} we show a sketch of the magnetic
fields of Abelian and non-Abelian kinds, respectively, in the case of
$\kappa>\mu$ (a) and in the case of $\kappa<\mu$ (b). The integral
over the plane of the Abelian magnetic field density is proportional
to the topological charge of the vortex, the winding number which in
turn renders the soliton topologically stable. The vortex solution
with negative winding number $k<0$ can be interpreted as an
anti-vortex. Hence, one could wonder which interpretation to give the
small substructure found in this vortex solution -- a small
anti-vortex trapped in the non-Abelian vortex, as a bound state, not
rendering the solution unstable. 

\bigskip

\noindent\emph{Opposite signs of coupling constants}\\
We will now consider taking one of the couplings to be negative, say
$\kappa<0$ and $\mu>0$. Choosing both signs negative yields the same
solution as already mentioned, however with flipped electric
fields. In the case of $\kappa>0$ and $\mu<0$, the solutions are
equivalent to the ones we will consider now, just with the signs
flipped of the electric fields. 
\begin{figure}[!tbp]
\begin{center}
\mbox{
\subfigure[]{\resizebox{!}{5.2cm}{\includegraphics{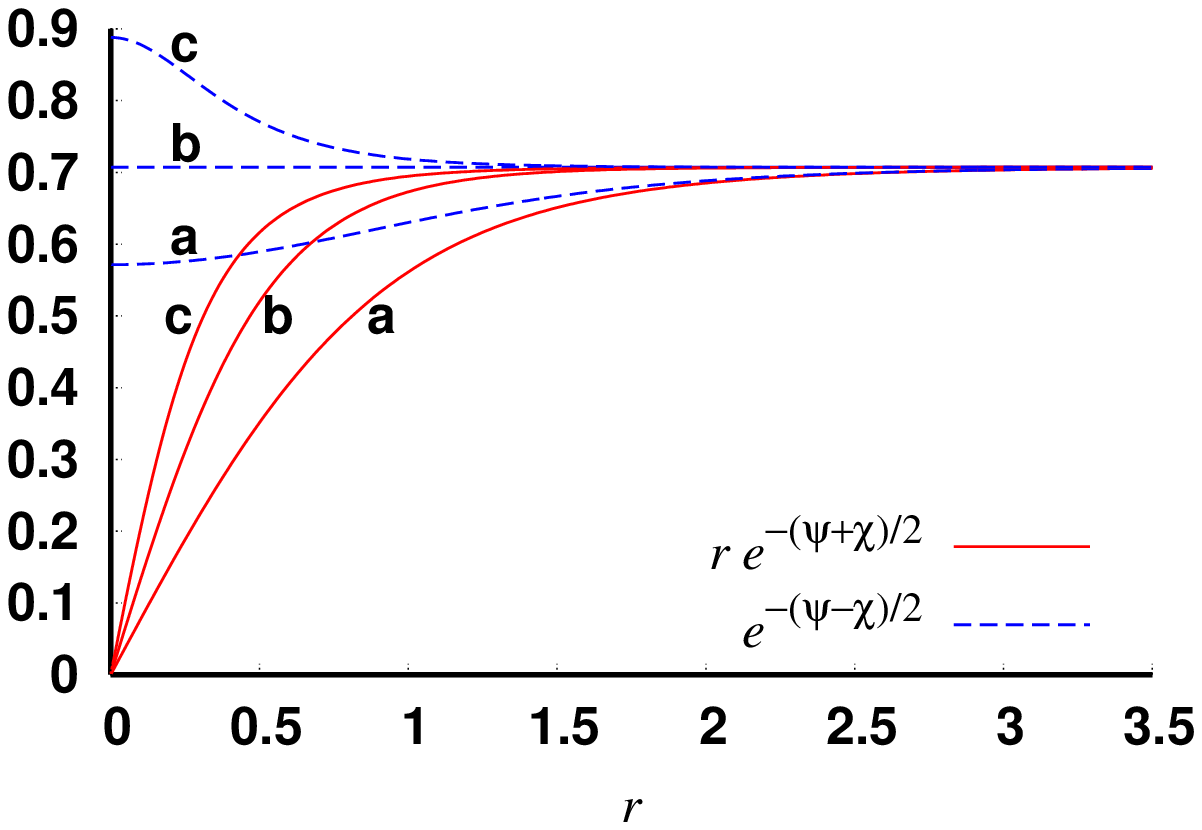}}}\quad
\subfigure[]{\resizebox{!}{5.2cm}{\includegraphics{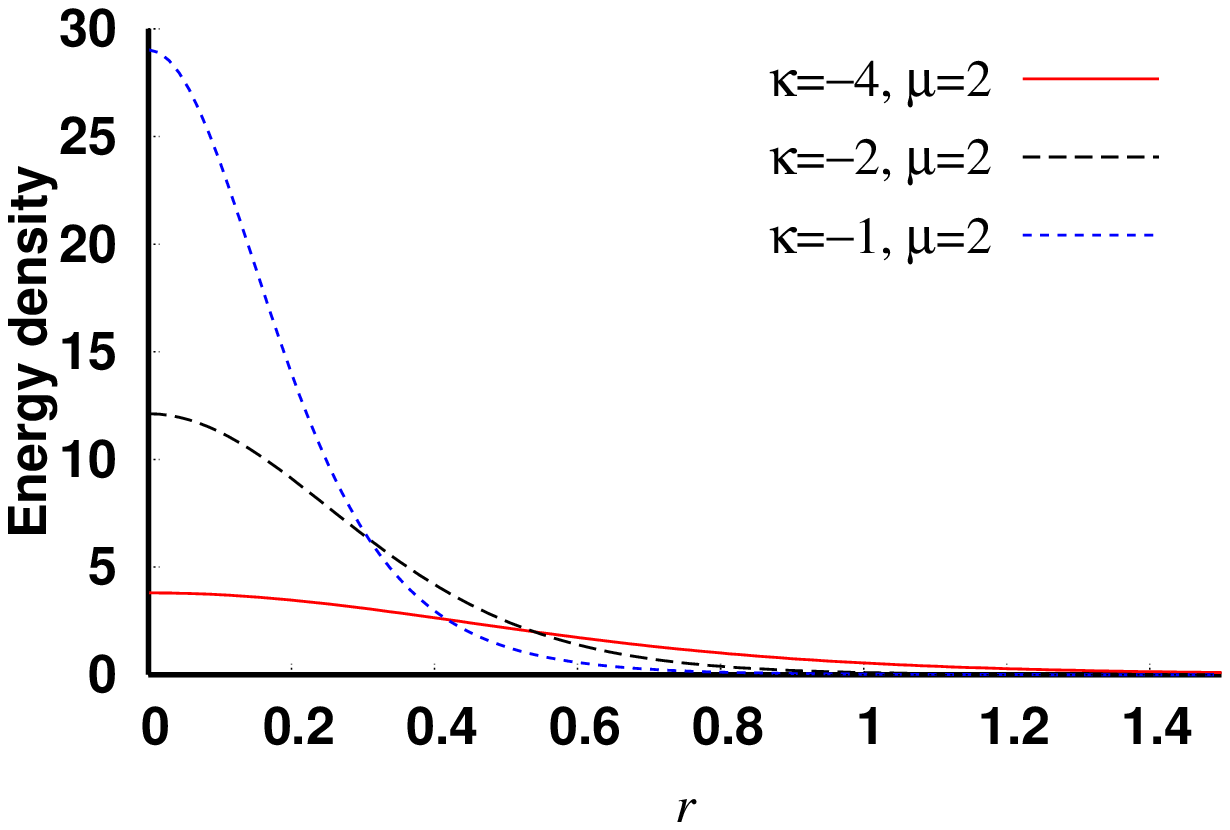}}}}
\caption{(a) Profile functions for three different values of the
  coupling constants; a: $\kappa=-4,\mu=2$; b: $\kappa=-2,\mu=2$; c:
  $\kappa=-1,\mu=2$; the functions are plotted in traditional style
  with the winding field rising linearly and the non-winding field
  being constant at the origin. The FI parameter $\xi=2$. Notice
  that the VEV for these functions is $2^{-\frac{1}{2}}$. (b) The
  energy density $\mathcal{E}$ for the vortex for the same three
  different values of the couplings with opposite signs. All the
  energy densities 
  integrate to $\pi\xi$, within an accuracy better than $\sim
  10^{-4}$. Notice that the extrema of the energy density is \emph{at
  the origin}, just as  in the case of the ANO vortices or the
  non-Abelian generalizations. } 
\label{fig:ns_profiles+en}
\end{center}
\end{figure}
\begin{figure}[!tbp]
\begin{center}
\mbox{
\subfigure[]{\resizebox{!}{5.2cm}{\includegraphics{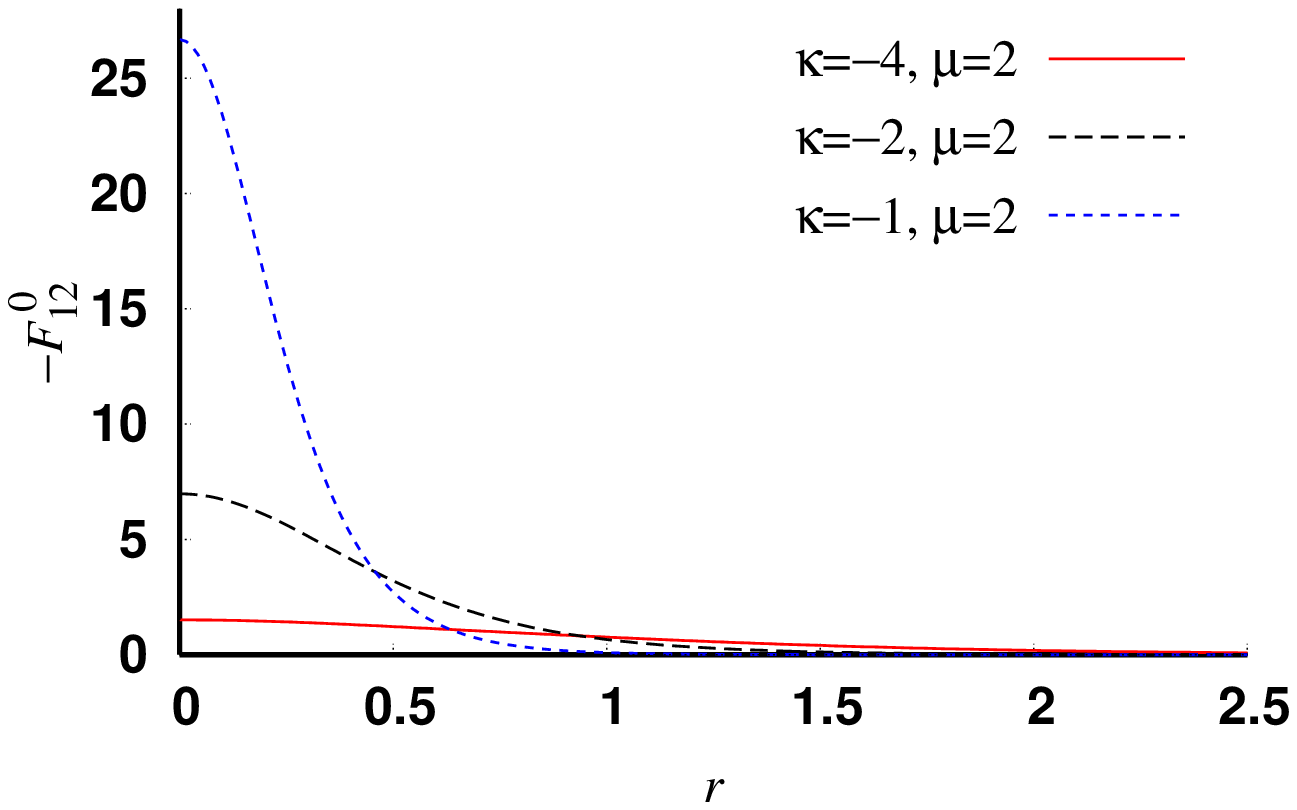}}}\quad
\subfigure[]{\resizebox{!}{5.2cm}{\includegraphics{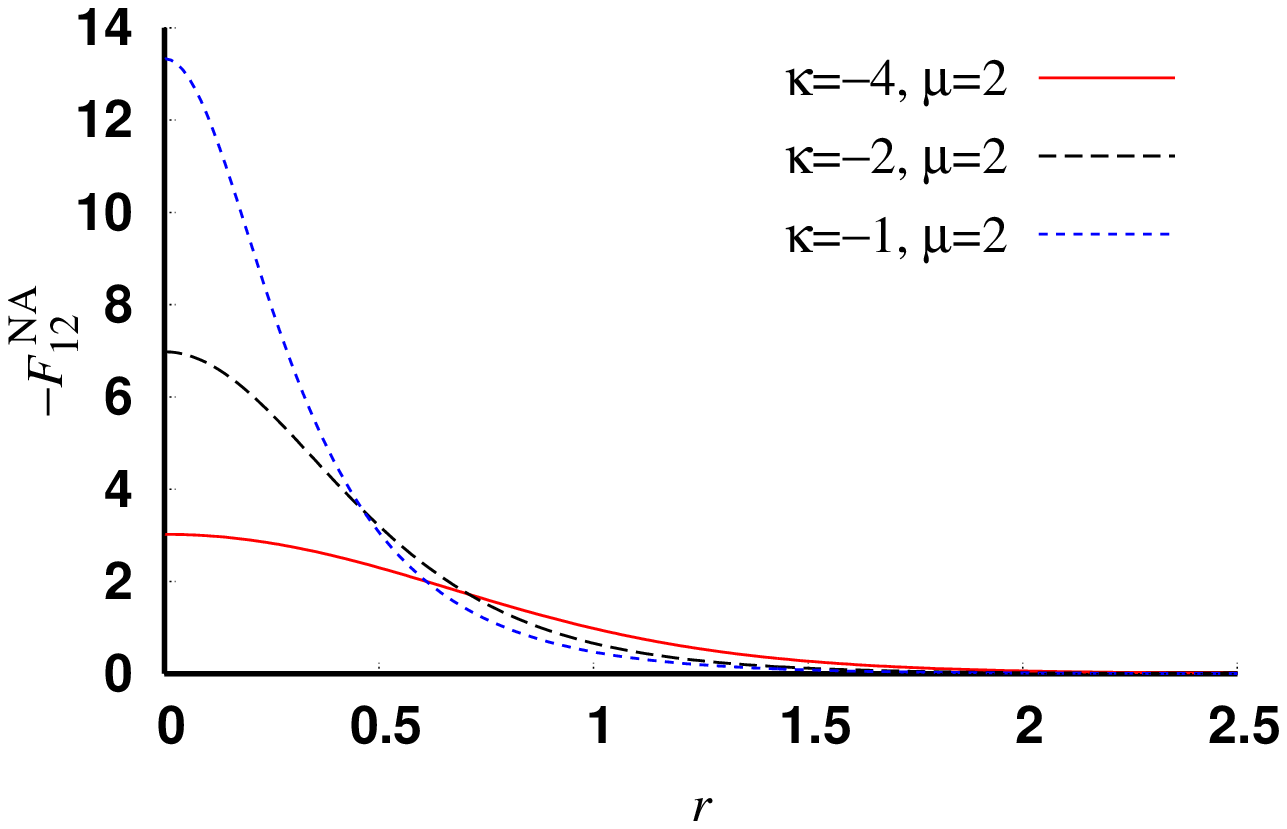}}}}
\caption{(a) The Abelian magnetic field $F_{12}^0$ (trace-part) and
  (b) the non-Abelian magnetic field $F_{12}^{\rm NA}$ (traceless part) for
  three different values of the couplings with opposite signs. Notice
  that the magnetic field density resembles that of the ANO vortex or
  the non-Abelian generalizations, viz.~they have the extrema at the
  origin. The FI parameter $\xi=2$. } 
\label{fig:ns_magflux}
\end{center}
\end{figure}
\begin{figure}[!tbp]
\begin{center}
\mbox{
\subfigure[]{\resizebox{!}{5.2cm}{\includegraphics{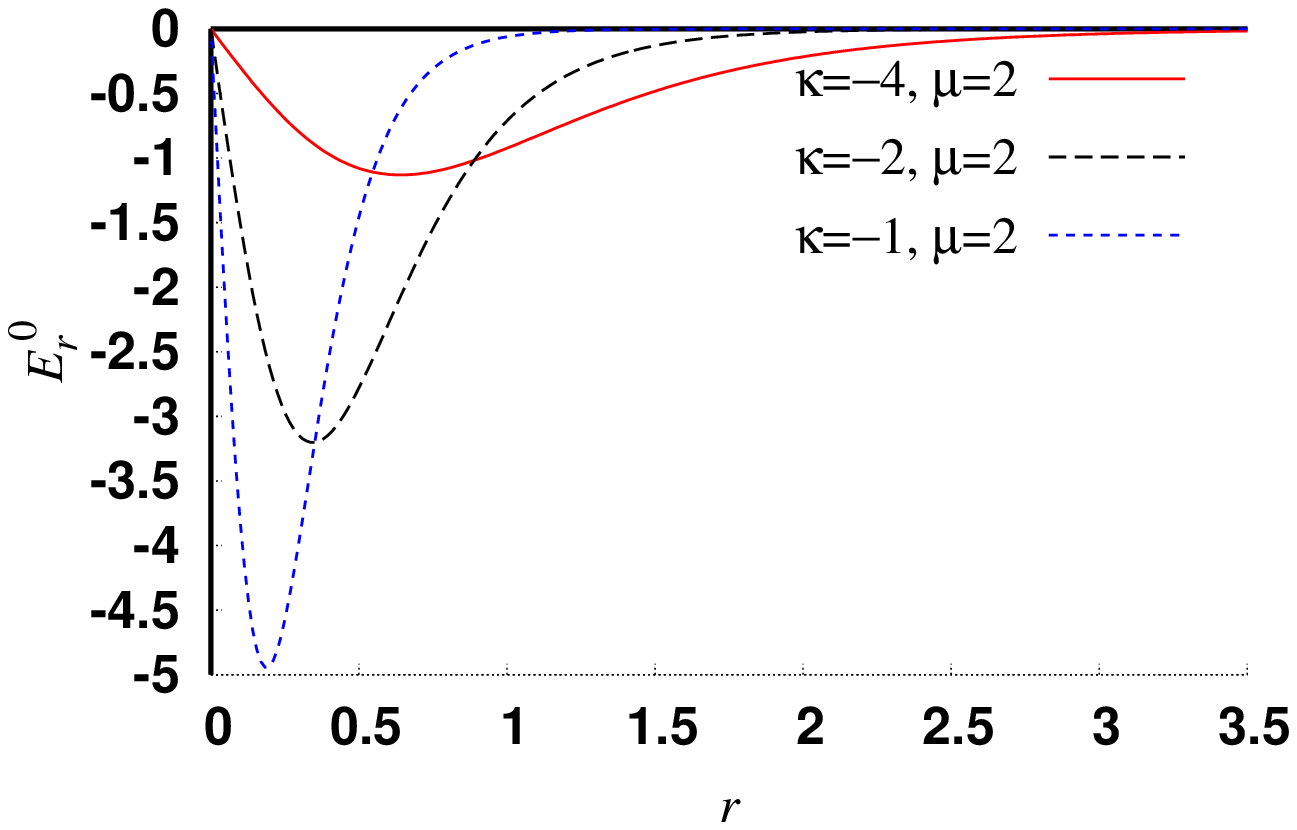}}}\quad
\subfigure[]{\resizebox{!}{5.2cm}{\includegraphics{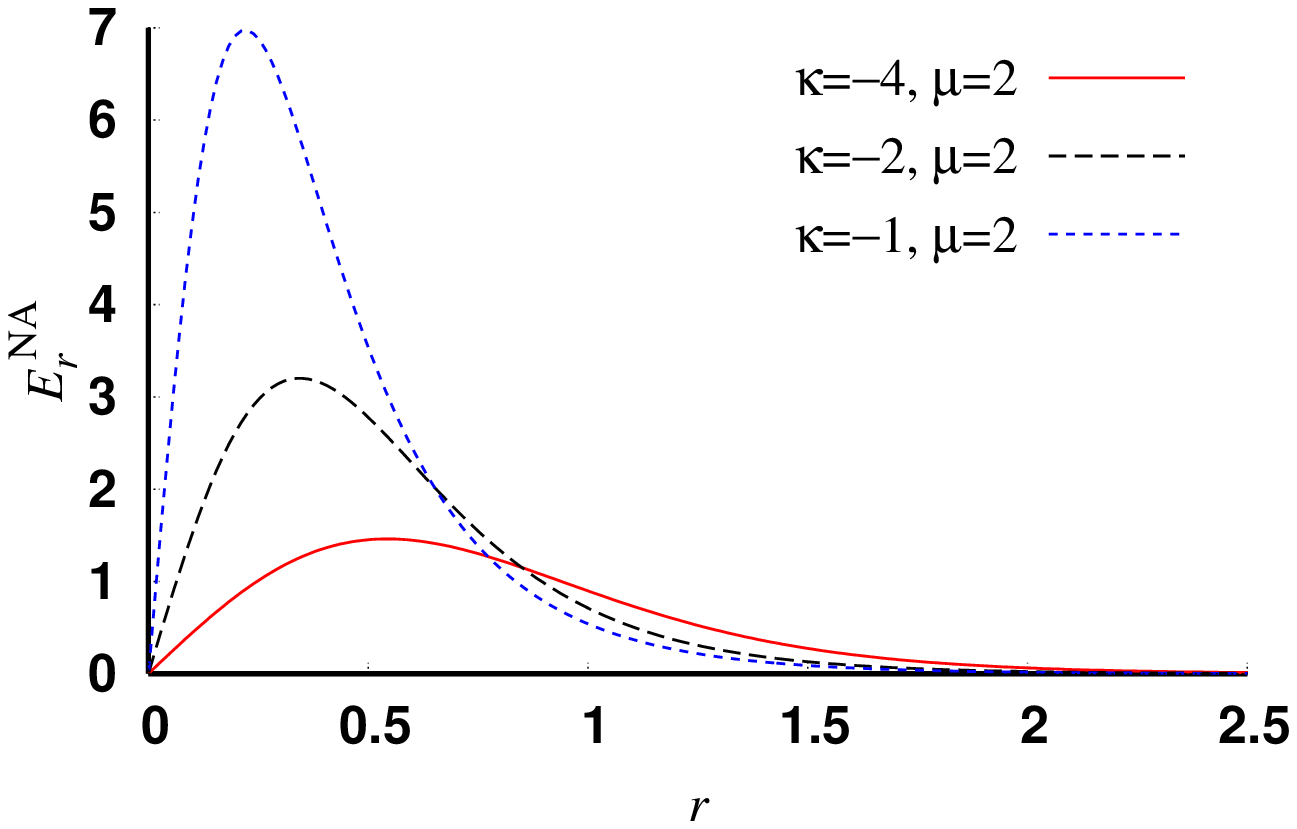}}}}
\caption{(a) The Abelian electric field in the radial direction
  $E_{r}$ (trace-part) and (b) The non-Abelian electric field
  $E_{r}^{\rm NA}$ (traceless part) for three different values of the
  couplings with opposite signs. The FI parameter $\xi=2$. Note that
  the electric fields are back-to-back. } 
\label{fig:ns_elecflux}
\end{center}
\end{figure}
The Chern-Simons characteristics have been lost in this case, the
vortex instead has the magnetic field concentrated at the origin --
just as in the case of the ANO vortex or the single $U(N)$ non-Abelian 
generalization. 
In Fig.~\ref{fig:ns_profiles+en} the profile functions and energy
densities for different solution are shown. In
Fig.~\ref{fig:ns_magflux} the corresponding magnetic fields are shown
while in Fig.~\ref{fig:ns_elecflux} the electric fields are shown.

\section{Discussion}

We have thus brought the powerful moduli matrix formalism into
the non-Abelian Chern-Simons model (which supports topological
non-Abelian vortices), and have conjectured that the moduli spaces
of the non-Abelian vortex solutions of these systems are indeed
identical to those of the vortex solutions
in the Yang-Mills-Higgs models with corresponding gauge groups. We
have not proved that every moduli matrix has a unique and existing
solution to the master equations found. Nevertheless we have argued
the plausibility of such a claim by taking the weak coupling limit which immediately
yields the lumps of the Yang-Mills-Higgs models, as it is just the
algebraic solutions to the $D$ term conditions. 

We have then studied some numerical solutions of non-Abelian vortices,
by choosing an Ansatz to the master equations, working mainly with the
$G'=SO(4)$ and $G'=USp(4)$ gauge groups. We have studied the case of
different couplings with both couplings positive yielding vortex
solutions with a small negative Abelian (non-Abelian) magnetic field
density at the origin and a corresponding positive non-Abelian
(Abelian) magnetic field density, which have a combination that is
always zero (at the origin). Keeping the couplings equal provides the
typical 
Chern-Simons characteristic that the magnetic field vanishes at the
origin yielding a ring structure. These new type of solutions
could perhaps be interpreted as an anti-vortex sitting inside the
non-Abelian vortex as a stable bound state, with the stability
provided by topological arguments. 

An interesting question is to which extent this substructure found in
the non-Abelian vortex solutions alters the dynamics of the vortices.

Furthermore, by changing the
relative sign of the coupling constants a vortex solution with the
magnetic field density concentrated at the origin has been found. 

An obvious future study related to these vortices and also to the ones
of Ref.~\cite{Eto:2009bg} could be to make an explicit construction
with exceptional groups and investigating the corresponding moduli
spaces. Especially interesting would be the center-less groups.

Another interesting path to follow is to consider the construction of
the non-Abelian vortices in Chern-Simons models with more
supersymmetries, e.g.~considering the model of
Aharony-Bergman-Jafferis-Maldacena \cite{Aharony:2008ug}. 
An Abelian non-relativistic Jackiw-Pi vortex
has already been found in this model \cite{Kawai:2009rc}. Another
attempt to construct vortices in the latter model has recently been
made, resulting in the non-Abelian vortex equations of the
Yang-Mills-Higgs models \cite{Kim:2009ny}.

\subsubsection*{Acknowledgements}

SBG thanks Minoru Eto, Jarah Evslin, Matteo Giordano, Kenichi Konishi,
Muneto Nitta, Giacomo Marmorini and Walter Vinci for fruitful
discussions.

\end{document}